\documentclass[11pt,preprint]{aastex}
\usepackage{graphicx}

\begin{document}

\newcommand{\Blos}{B_{los}}
\newcommand{\Btot}{B_{tot}}
\newcommand{\Bobs}{B_{los}}
\newcommand{\Bpar}{B_{los}}
\newcommand{\Nobs}{N_{los}}
\newcommand{\Nperp}{N_{\perp}}
\newcommand{\Va}{V_{turb}}
\newcommand{\Vntobs}{V_{turb, los}}

\title{THE MILLENNIUM ARECIBO 21-CM ABSORPTION LINE SURVEY. IV. \\
STATISTICS OF MAGNETIC FIELD, COLUMN DENSITY, AND TURBULENCE \\
\today
}

\author{Carl Heiles}
\affil{Astronomy Department, University of California,
    Berkeley, CA 94720-3411; cheiles@astron.berkeley.edu}

\author{T.H. Troland}
\affil{Department of Physics and Astronomy, University of Kentucky,
Lexington, KY; troland@pa.uky.edu}

\begin{abstract}

	We discuss observations of the magnetic field, column density,
and turbulence in the Cold Neutral Medium (CNM). The observed quantities
are only indirectly related to the intrinsic astronomical ones. We
relate the observed and intrinsic quantities by relating their
univariate and bivariate probability distribution functions (pdfs). We
find that observations of the line-of-sight component of magnetic field
do not constrain the pdf of the total field $B_{tot}$ very well, but do
constrain the median value of $B_{tot}$. 

	In the CNM, we find a well-defined median magnetic field $6.0
\pm 1.8$ $\mu$G. The CNM magnetic field dominates thermal motions.
Turbulence and magnetism are in approximate equipartition. We find the
probability distribution of column density $\Nperp (HI)$ in the sheets
closely follows $\Nperp (HI)^{-1}$ over a range of two orders of
magnitude, $0.026 \lesssim \Nperp (HI) \lesssim 2.6 \times 10^{20}$
cm$^{-2}$. The bivariate distributions are not well enough determined to
constrain structural models of CNM sheets.

\end{abstract}

\tableofcontents

\section{INTRODUCTION} \label{introduction}

Beginning February 1999 we used the Arecibo telescope \footnote{The
Arecibo Observatory is part of the National Astronomy and Ionosphere
Center, which is operated by Cornell University under a cooperative
agreement with the National Science Foundation.} to begin a
series of Zeeman-splitting measurements of the 21-cm line in absorption
against continuum radio sources.  Heiles \& Troland (2004; Paper III and
references therein)  discussed technical details of the observing
technique and data reduction and presented observational results,
including magnetic field measurements.  As described in Paper III, our
analysis of the data identified Gaussian components in the HI absorption
spectra.  We assume  that each component samples a single, isothermal,
sheet-like region of CNM.  For 69 such components, we are able to
estimate (1) the line-of-sight magnetic field strength $\Bobs$, often
subject to significant error, (2) the line-of-sight column density
$N(HI)_{los}$, and (3) the contribution of turbulence $V_{turb,los}$ to
the line-of-sight velocity dispersion.  Systematic instrumental errors
are small and the uncertainties are Gaussian distributed. Therefore, 
our survey has yielded a statistically well-defined ensemble of observed
values for each of these three CNM quantities.  

	From these data, we seek to answer several astrophysically
significant questions.  For example, what are the probability
distribution functions (pdfs) of magnetic field strengths, column
densities and turbulent energies in the CNM?  Are any of these
quantities statistically related to each other?  If so, can we determine
if the magnetic field lies preferentially in the planes of CNM sheets or
perpendicular to them?  What is the ratio of magnetic to turbulent
energy in the CNM?  As we will see below, the data provide
answers to some, but not all, of these questions.

	We observe quantities that are only indirectly related to the
intrinsic astronomical ones. For the magnetic field, we observe the line
of sight component $\Bobs$, not the total field $\Btot$. For the column
density $N(HI)$, we must account for the fact that the Cold Neutral
Medium is in sheets, not spheres; our observed column densities $\Nobs$
are always larger than the intrinsic column density normal to the sheet
$\Nperp$. To properly interpret our results and to address the questions
listed above, we must consider how the intrinsic and observed quantities
are related in a statistical sample. These statistical transformations,
and the results of applying them to our data, are the focus of the
current paper. 

	We treat both univariate distributions and bivariate
distributions. The univariate distributions are most interesting because
we obtain definitive results; in contrast, the possible correlations
that could be revealed by bivariate distributions are obscured by noise
and inadequate numbers of data. Accordingly, the reader who is
interested in astrophysical results can concentrate on the sections
dealing with univariate distributions of magnetic field, column density,
and turbulent velocity. We recommend beginning with \S \ref{important},
which introduces the notation and concepts of intrinsic and observed
quantities. We develop the theoretical relationships between observed
and intrinsic univariate distributions in \S \ref{thconv0}, and use
these to obtain the actual intrinsic univariate distributions in \S
\ref{thconvuni}. Finally, we discuss astrophysical implications in \S
\ref{magdiscussion}, \S \ref{ndiscussion}, \S \ref{vdiscussion}, and
\S \ref{finalcomments}.

\section{SOME IMPORTANT WORDS ON NOTATION} \label{important}

	We need to introduce important distinguishing nomenclature
because of the anisotropic nature of magnetic-field-related quantities.
First, we make the crucial distinction between {\it line-of-sight} ($los$) 
quantities, such as $\Blos$, and the {\it intrinsic astronomical} ones
such as $B_{tot}$. We will often refer to the latter with one the shorter
terms ``intrinsic'' or ``astronomical''. The important point is to
distinguish between observed and intrinsic quantities. 

	The line-of-sight component is the quantity to which the
telescope responds, and we designate these with the subscript $los$. 
There is an additional complication for the magnetic field, which is the
presence of significant instrumental noise. Thus, we must distinguish
between the line-of-sight component of the field, designated by
$B_{los}$, and the actual measured value, designated by $B_{obs}$. We
define 

\begin{mathletters}
\begin{equation}
B_{los} \equiv the \ line{\rm -}of{\rm -}sight \ component \ of \
B_{tot} \ ;
\end{equation}
\begin{equation}
B_{obs} \equiv the \ observed \ field \ strength  \ .
\end{equation}

\noindent The essential difference between $B_{los}$ and $B_{obs}$ is
\begin{equation}
B_{obs} = B_{los} + \delta B_{noise}
\end{equation}
\end{mathletters}
\noindent where $\delta B_{noise}$ is the uncertainty contributed by
random measurement noise.

	Noise is small enough to neglect for column density and
velocity, so the observed quantities are essentially identical to the
$los$ ones; we will use the subscript $los$ for the observed quantities,
with the implicit assumption that their uncertainties from noise are
negligible. For column density, we always refer to HI so we often write
$N$ instead of $N(HI)$. We usually consider sheets, for which the column
density perpendicular to the face is $N(HI)_\perp$ and for which the
apparent column density along the line of sight is

\begin{equation}
N_{los} \equiv N(HI)_{los} = N(HI)_{obs} = {N(HI)_\perp \over \cos
\theta} \ ,
\end{equation}

\noindent where $\theta$ is the angle between the sheet's normal vector
and the line of sight. Similarly, for velocity linewidths, we use the
symbol $V_{turb,los}$ to indicate the line-of-sight component of the
turbulent velocity. 

	Finally, the probability density functions (pdf) of intrinsic
quantities, such as $B_{tot}$, differ from those of the line-of-sight
quantities. We will always use the symbol $\phi$ for the pdfs of
intrinsic quantities, and the symbol $\psi$ for line-of-sight ($los$) or
the observed ones.  Unless otherwise specified, units are always as   
follows: magnetic field $B$ in $\mu$Gauss; column density $N$ in HI 
atoms $10^{20}$ cm$^{-2}$; velocity $V$ in km s$^{-1}$.

\section{THEORETICAL CONVERSIONS OF INTRINSIC ASTRONOMICAL PROBABILITY
DENSITY FUNCTIONS TO OBSERVED ONES} \label{thconv0}

	We begin with some light theory by considering elementary
transformations of magnetic field and column density in statistical
distributions. First, we consider how the intrinsic probability density
function (pdf) of $\Btot$, defined as $\phi(\Btot)$, converts to the
observed histogram or pdf of observed $\Bpar$, defined as $\psi(\Bobs)$,
under the assumption that fields are randomly oriented with respect to
the observer. Note the important distinction: $\Btot$ is the total field
strength; $\Bobs$ is the observed line-of-sight component, which is
always smaller.

	Next we incorporate one of the fundamental results of Heiles \&
Troland (2003; Paper II), namely that the CNM components are thin
sheets.  We define $N_\perp = N(HI)_\perp$ as the HI column density
perpendicular to the sheet; the observed quantity is
$N_{los}=N(HI)_{los}$, which is always larger.  As with the magnetic
field, we consider how the pdf $\phi(\Nperp)$ converts to the observed
histogram or pdf $\psi(\Nobs)$, again under the assumption that the
sheets are randomly oriented with respect to the observer.  We also
consider the statistical transformation of the distribution of the
actual nonthermal velocity dispersion $\Va$ to that of the observed one
$\Vntobs$ under the assumption that turbulence is perpendicular to the
magnetic field. 

	We then tackle the bivariate distributions. First we consider
how the bivariate distribution $\phi(\Btot, \Nperp)$ converts to the
observed one $\psi(\Bobs, \Nobs)$. We then assume two extreme models,
one with the fields always perpendicular to the sheets and one with
fields parallel to the sheets, and assuming random orientations derive
the observed bivariate distributions $\psi(\Bobs, \Nobs)$ for the two
cases. We illustrate and discuss the result by considering
delta-function distributions of $\Btot$ and $\Nperp$, and we also apply
the transformation of the observed $\psi(\Bobs, \Nobs)$ to its intrinsic
counterpart $\phi(\Btot,\Nperp)$. Finally, we examine the bivariate
distributions involving the pairs $(\Vntobs, \Nobs)$ and $(\Vntobs,
\Bobs)$, which produces little in the way of useful results. 

\subsection{Conversion of the univariate distributions} \label{thconvuni}

\subsubsection{Conversion of the intrinsic \boldmath{$\phi(\Btot)$} to
the observed \boldmath{$\psi(\Bobs)$} } \label{univar_B}

	We first consider the simple case in which all clouds have the
same $\Btot$, which is randomly oriented with respect to the
observer. The line of sight component $\Bobs$ is 

\begin{equation}
\Bobs = \Btot \cos \theta \ ,
\end{equation}

\noindent where $\theta$ is the angle between the field direction and
the line of sight. $\theta$ can run from 0 to $\pi$, but the intervals
from 0 to $\pi \over 2$ and $\pi \over 2$ to $\pi$ are identical except
for a change of sign in $\Bobs$. It's simpler and no less general to
consider the smaller interval $\theta$ from 0 to $\pi \over 2$ so that
we can ignore the slight complications of the sign change. In this case,
the pdf of $\theta$ is  the familiar

\begin{equation}
\phi_\theta(\theta) = \sin \theta 
\end{equation}

\noindent and we wish to know the pdf of $\Bobs$, which is given by
(see Trumpler and Weaver 1953 for a discussion of these conversions)

\begin{equation}
\psi({\Bobs}) = \phi_\theta[ \theta(\Bobs)] \left| 
	{d[\theta(\Bobs)] \over d\Bobs} \right|  \ ,
\end{equation}

\noindent which gives

\begin{eqnarray} \label{Bobs}
\psi({\Bobs}) = 
\left\{ 
\begin{array}{ll}
{1 \over \Btot} & {\rm if} \  0  \leq \Bobs \leq \Btot \\ 
        0       & {\rm otherwise}
\end{array}
\right. 
\end{eqnarray}

\noindent In other words, $\Bobs$ is uniformly distributed between the
maximum possible extremes 0 and $\Btot$ (actually $\pm\Btot$). This
leads to the well-known results that in a large statistical sample for
which a constant $B_{tot}$ is viewed at random angles, both the median
and the mean observed field strengths are half the total field strength
and that $\Bpar^2 = {\Btot^2 \over 3}$. More generally, observed fields
are always smaller than the actual total fields, and with significant
probability they range all the way down to zero.

	Now suppose $\Btot$ has an arbitrary pdf $\phi(\Btot)$. The
univariate pdf $\phi(\theta)$ becomes the bivariate pdf
$\phi(\Btot,\theta)$, and we assume $\Btot$ is independent of the
observer's location so that $\phi(\Btot,\theta)=
\phi_{\Btot}(\Btot)\phi_\theta(\theta)$. Note that we introduce
subscripts on the different $\phi$'s to distinguish them, instead of
designating them with different Greek letters. To obtain  $\psi(\Bobs)$
we again follow the standard techniques; it's easy to integrate over
$\theta$ and obtain

\begin{equation} \label{Bpsiphi}
\psi(\Bpar) = \int_{[{\Bobs}>{\Btot}_{min}]}^\infty {\phi(\Btot) 
	\over \Btot} d\Btot \ , 
\end{equation}

\noindent where the symbol $[{\Bobs}>{\Btot}_{min}]$ means the larger of
the two quantities. The presence of $\Btot$ in the denominator means
that smaller ranges of $\Bobs$ are emphasized. This is commensurate
with the equation \ref{Bobs}'s uniform pdf for a single field
value. We note that this can be regarded as an integral equation for
$\phi(\Btot)$ and it is straightforward to invert.

\begin{figure}[h!]
\begin{center}
\includegraphics[width=3.0in] {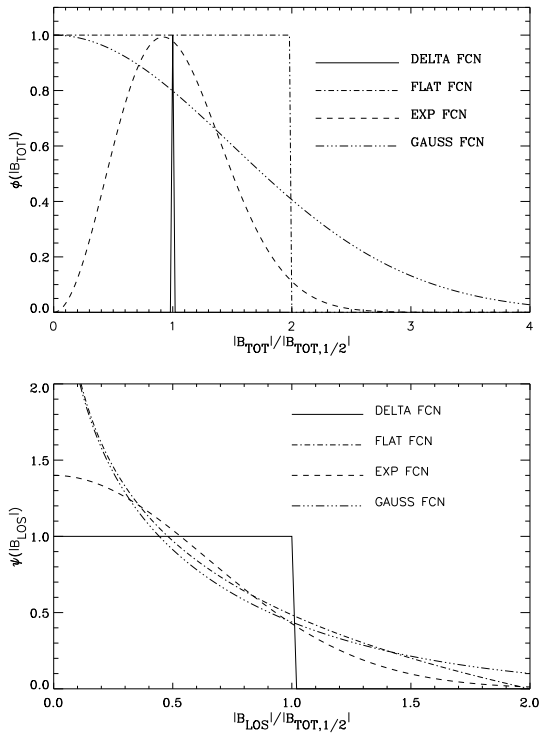}
\end{center}

\caption{Top panel: The intrinsic $\phi(B_{tot})$ for four
representative functional forms. Bottom panel: their line-of-sight
counterparts $\psi(B_{los})$. The vertical scales are arbitrary.
\label{pdf_fig0} } \end{figure}

	Figure \ref{pdf_fig0} illustrates the solution of equation
\ref{Bpsiphi} for four functional forms of $\phi(B_{tot})$ plotted
against $|B| \over |B_{1/2}|$, where the subscript $1/2$ denotes the
median value. These forms include the following:
\begin{enumerate}

\item $\phi$ a Kronecker delta function (DELTA FCN),
$\phi_{\Btot}(\Btot) = \delta(\Btot - {\Btot}_{,0})$, yielding $\psi$ a
flat function (as discussed immediately above, equation \ref{Bobs});

\item $\phi$ a flat distribution (FLAT FCN) between $0 \le |B_{tot}| \le
B_0$, yielding $\psi \propto \ln\left({ B_0 \over B_{los}}\right)$;

\item  $\phi$ a weighted Gaussian (EXP FCN),

\begin{equation} \label{Btotexp}
\phi({\Btot}) = {{\sqrt{2 \over \pi B_0^2}}} \ {\Btot ^2 \over 2B_0^2}
\  e^{-( \Btot ^2 / 2 B_0^2)} \ ,
\end{equation}

\noindent yielding $\psi$ a Gaussian with dispersion $B_0$.

\item $\phi$ a Gaussian (GAUSS FCN) with dispersion $B_0$, yielding
$\psi \propto E_1 \left(B_{los}^2 \over 2B_0^2 \right)$, where $E_1$ is
the exponential integral of order 1. 

\end{enumerate}

	All four $\phi(B_{tot})$ are plotted with respect to $B_{tot}
\over B_{tot,1/2}$, so the medians of all lie at unity on the $x$-axis.
However, the means differ. Similarly, the medians and means of the
associated $\psi(B_{los})$ differ from each other. These relationships
between median and mean are summarized in Table \ref{medianmean}. The
medians and means for $\psi(B_{los})$ are all about half those for
$\langle B_{tot} \rangle$, which is a direct result of the weighting by
$B_{tot}^{-1}$ in equation \ref{Bpsiphi}.

	Figure \ref{pdf_fig0} is disappointing from the observer's
standpoint, because the observed distributions $\psi(B_{los})$ do not
differ very much. These differences become smaller---inconsequential, in
fact---when one includes measurement noise, as we shall discuss in
detail for our data in \S \ref{Btotderivation}. Unfortunately, given the
inevitable errors in {\it any} observation that is sensitive to
$B_{los}$, it seems practically impossible to distinguish among
different functional forms for $\phi(B_{tot})$. Nevertheless, the
average value of $B_{los}$ is close to half the average value of
$B_{tot}$ for a wide range of intrinsic pdfs of the latter; this also
applies to the medians, but less accurately. Therefore, this rule of
thumb may be used to estimate the median or average $B_{tot}$ from an
ensemble of measurements of $B_{los}$. 

\begin{deluxetable}{ccccc}  
\tablewidth{0pc}
\tablecaption{Medians and Means of $B_{tot}$ and $B_{los}$ for Representative
pdfs \label{medianmean} }
\tablehead{
\colhead{ $\phi(B_{tot}$ }&
\colhead{ $B_{tot,1/2}$ }&
\colhead{ $\langle B_{tot} \rangle$ } &
\colhead{ $B_{los,1/2}$ }&
\colhead{ $\langle B_{los} \rangle$ }}
\startdata
DELTA FCN & 1.00 & 1.00  & 0.50  & 0.50  \\
FLAT FCN & 1.00  & 1.00  & 0.40  & 0.52  \\
GAUSS FCN & 1.00 & 1.18  & 0.38  & 0.59  \\
EXP FCN   & 1.00 & 1.04  & 0.44  & 0.51  \enddata
\end{deluxetable}

\subsubsection{Conversion of the intrinsic \boldmath{$\phi(\Nperp)$} to
the observed \boldmath{$\psi(\Nobs)$} for sheets} \label{univarN}

	Here we assume the CNM is distributed in sheets having HI column
density $\Nperp$ in the direction perpendicular to the sheet. If the
normal vector to the sheet is oriented at angle $\theta$ with respect to
the line of sight, we have

\begin{equation}
\Nobs = { \Nperp \over \cos\theta} \ .
\end{equation}

\noindent To find $\psi(\Nobs)$ we follow the same procedures as above
in \S \ref{univar_B}. For a single value of $\Nperp$ we obtain

\begin{eqnarray} \label{Nobs}
\psi(\Nobs)= 
\left\{ 
\begin{array}{ll}
{\Nperp \over \Nobs^2} & {\rm if} \  \Nobs \geq \Nperp \\ 
        0       & {\rm otherwise}
\end{array}
\right. 
\end{eqnarray}

\noindent For a single $\Nperp$, $\Nobs$ has a long tail extending to
infinity. The median value of $\Nobs$ is $N_{los, 1/2} = 2 \Nperp$,
reflecting the increased observed column for tilted sheets. The mean
value of $\Nobs$ ($\langle N_{los} \rangle$) is not defined because the
integral diverges logarithmically; of course, this doesn't occur in the
real world, where sheets don't extend to infinity. For example, if all
sheets have aspect ratio $5:1$, then $\langle N_{los} \rangle = 1.6
\Nperp$ and $N_{los, 1/2} = 1.7 \Nperp$. 

	For an arbitrary pdf $\phi(\Nperp)$ and infinite slabs,  we
obtain

\begin{equation}  \label{Npsiphi}
\psi(\Nobs )= {1 \over \Nobs ^2} \int_0^{[\Nobs < {\Nperp}_{max}]} 
	\Nperp \ \phi(\Nperp) \ d\Nperp \ .
\end{equation}

\noindent As above, this can be regarded as an integral equation for
$\phi(\Nperp)$ and it is almost as straightforward to invert. 

\subsubsection{Conversion of the intrinsic \boldmath{$\phi(\Va)$} to the
observed \boldmath{$\psi(\Vntobs)$} for turbulence perpendicular to
$\Btot$} \label{univarV}

	The observed velocity width comes from two sources, thermal and
nonthermal. We can separate these because we have independent
measurements of the kinetic temperature. Thus, we can derive the
line-of-sight (los) nonthermal (``turbulent'') line width and its
associated energy density. How this relates to the total turbulent
energy density depends on whether the turbulence is $1d$, $2d$, or $3d$.

        Here we assume that this turbulence is $2d$, i.e.\ we assume  
that it is restricted to motions perpendicular to the mean magnetic
field. We make this assumption because we show below that the typical
turbulent Mach number is equal to 3.7.	For velocities that are parallel
to the magnetic field, such turbulence would produce strong shocks that
would damp very rapidly. However, velocities that are perpendicular to
the magnetic field can be as high as the Alfv\'en velocity without
producing shocks; this is the basis for considering turbulent motions
perpendicular to the mean field. We caution, however, that numerical
simulations of magnetohydrodynamical turbulence find that magnetic
fields do not ameliorate turbulent dissipation (MacLow et al.\ 1998), so
our assumption might not have any basis in physical reality; in this
case the turbulence is isotropic ($3d$) and the observed distribution
$\psi(\Vntobs)$ is equal to the intrinsic one $\phi(\Va)$.	

	Suppose that the magnetic field is oriented at angle $\theta$
with respect to the line of sight, as in \S \ref{univar_B}. Suppose that
turbulent motions are perpendicular to the fieldlines and isotropic in
the azimuthal directions around the fieldlines, and along one direction
perpendicular to $B$ have width $\Va$; this makes the full turbulent
width $2^{1/2} \Va$. Then line-of-sight width is 

\begin{equation}
\Vntobs =\Va \sin \theta \ .
\end{equation}

\noindent To find $\psi(\Vntobs)$ we again follow the same procedures as
in \S \ref{univar_B} and obtain

\begin{eqnarray} \label{Vntobs}
\psi(\Vntobs)  =
\left\{ 
\begin{array}{ll}
{\Vntobs \over \Va} 
\left[ \Va^2 -  \Vntobs^2 \right]^{-1/2} &  
{\rm if} \  \Vntobs < \Va \\ 
        0       & {\rm otherwise}
\end{array}
\right. 
\end{eqnarray}

\noindent Given a value for  $\Va$, $\psi(\Vntobs) \rightarrow \infty$
as $\Va \rightarrow \Vntobs$, but the cumulative distribution is
well-defined with 

\begin{equation}
cum(\Vntobs) = 1 - \left[ 1 - \left( \Vntobs \over \Va \right) ^2
\right]^{1/2}
\end{equation}

\noindent which gives the median $V_{turb,los,1/2} = 0.87\Va$ median and
mean $\langle{\Vntobs}\rangle = 0.79\Va$. These high values reflect the
large fraction of sheets tilted to the line of sight, where $\Vntobs$ is
large.

	For an arbitrary pdf $\phi(\Va)$ we obtain

\begin{equation} \label{volterra1}
\psi(\Vntobs) = \int_{\Vntobs}^\infty
{\Vntobs \over \Va } 
\left[ \Va^2 - \Vntobs^2 \right]^{-1/2}  \phi_{\Va}(\Va )  d\Va
\end{equation}

\noindent In contrast to the two cases above, this integral equation is
not straightforward to invert. It is a Volterra equation of the first
kind and, following the identical example in Trumpler \& Weaver (1953),
it can be rewritten as Abel integral equation. The analytic solution is

\begin{equation} \label{abeleqn}
\phi(\Va)= -{2 \Va ^2 \over \pi} {d \over d\Va} \int_{\Va}^\infty
{\Va \over \Vntobs^2} \left[\Vntobs^2 - \Va^2\right]^{-1/2} 
\psi(\Vntobs) d\Vntobs 
\end{equation}

\subsection{Conversion of the intrinsic bivariate distribution
\boldmath{$\phi(\Btot, \Nperp$)} to the observed \boldmath{$\psi(\Bpar,
\Nobs)$}} \label{bivarBN}

	We have measured both $\Bobs$ and $\Nobs$ and wish to know the
bivariate distribution $\psi(\Bobs, \Nobs)$. We proceed by first
assuming that $\Btot$, $\Nperp$, and of course $\theta$ are all
uncorrelated. To proceed we consider two different models.

\subsubsection{Case of \boldmath{$\Btot$} perpendicular to the sheet}
\label{btotperp}

	We refer to this as the {\it perpendicular model}. The only angle
involved is $\theta$, so that  the original pdf is the trivariate
$\phi(\Btot,\Nperp,\theta)$.  This case is simplified because only $\cos
\theta$ is involved, which makes $\Bobs \propto {1 \over \Nobs}$. 
Converting the original distribution $\phi( \Btot, \Nperp, \theta)$ to
the one involving the observed parameters yields

\begin{equation} \label{perpsheet}
\psi( \Bobs, \Nobs, \Nperp) = {1 \over \Nobs} 
\phi_{Btot} \left( {\Bobs \Nobs \over \Nperp} \right) \phi_{\Nperp}(\Nperp)
\end{equation}

\noindent Here we have chosen to eliminate $\Btot$ and express the
result in terms of $\Nperp$; we could have gone the other way. To obtain
$\psi$ in terms of only the observed quantities, we need to integrate
over $\Nperp$, but we cannot do this without knowing
$\phi_{\Nperp}(\Nperp)$.  Later we use the one obtained from
observations. 

\begin{figure}[!p]
\begin{center}
\includegraphics[width=5.0in] {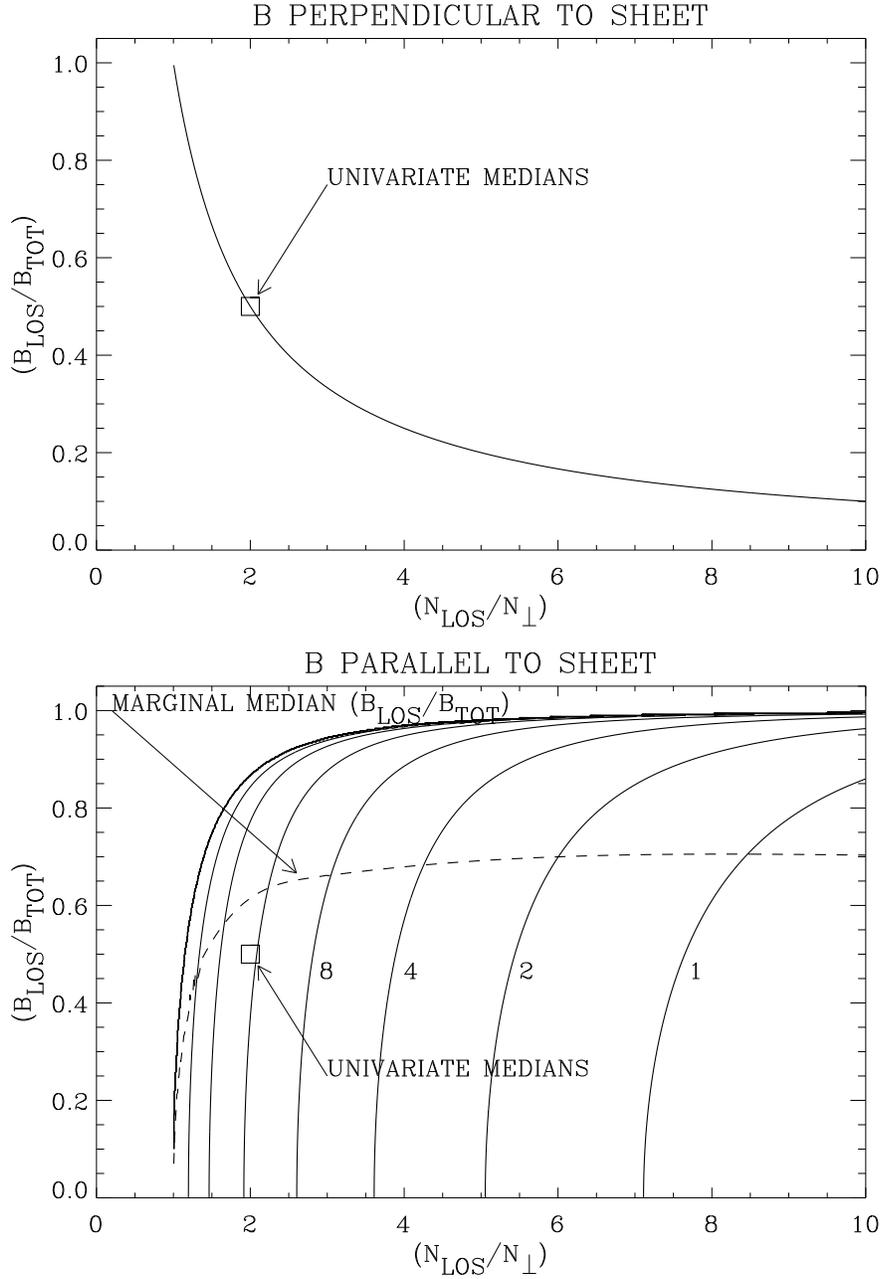}
\end{center}

\caption{The theoretical observed joint pdfs $\psi(\Bpar, \Nobs)$ for
the illustrative case of $\delta$-function distributions for $\Bpar$ and
$\Nobs$. The top panel shows the pdf for $\Btot$ perpendicular to the
sheets; it degenerates into a single line. The bottom panel is for
$\Btot$ parallel to the sheets; contours are spaced by factors of 2 with
arbitrary scaling, and the dashed line shows the median $\Bobs$ versus
$\Nobs$. \label{twodpdfplot} } \end{figure}

	For now we consider the illustrative case for which all
$B_{tot}$ and $N_\perp$ are identical, i.e.\ $\phi_{\Btot}(\Btot) =
\delta(\Btot - {\Btot}_{,0})$ and $\phi_{\Nperp}(\Nperp) = \delta(\Nperp
- {\Nperp}_0)$, where $\delta$ is the Kronecker delta function. This is
a trivial case because all observed points fall on the line

\begin{equation}
\Bobs= {\Btot}_{,0} {{\Nperp}_0 \over \Nobs}
\end{equation}

\noindent which is shown in the top panel of Figure \ref{twodpdfplot}.

\subsubsection{Case of \boldmath{$\Btot$} parallel to the sheet}
\label{Btotpar}

	This case is more complicated because $\Bpar$ depends on two
angles. These are $\theta$, which is the tilt of the sheet with respect
to the observer, and $\Phi$, which is the azimuthal angle of the field
within the sheet. We have $\Bpar = \Btot \sin \theta \sin \Phi$. This
means the pdf $\phi$ is quadrivariate. Again we assume everything is
uncorrelated and eliminate $\theta$ and $\Phi$. Obtaining the observed
distribution is somewhat cumbersome but yields the surprisingly
straightforward result

\begin{equation} \label{parsheet}
\psi(\Bpar, \Nobs, \Btot, \Nperp)=
{\Nperp \over \pi \Nobs}[ (\Btot\Nobs)^2 - (\Btot\Nperp)^2 -
(\Bpar\Nobs)^2]^{-1/2} 
\phi_{\Btot}(\Btot) \phi_{\Nperp}(\Nperp)
\end{equation}

	Later we use observed distributions, but for now we again
consider the illustrative case of delta functions for $\Btot$ and
$\Nperp$. Integrating over $\Btot$ and $\Nperp$ yields

\begin{equation}
\psi(\Bpar , \Nobs)=
{ {\Nperp}_0 \over \pi \Nobs} 
[ ({\Btot}_{,0} \Nobs)^2 - ({\Btot}_{,0} {\Nperp}_0)^2 -(\Bpar \Nobs)^2 ]^ {-1/2}
\end{equation}

\noindent which is illustrated in the bottom panel of Figure
\ref{twodpdfplot}.

\subsubsection {Discussion of Figure \ref{twodpdfplot}}
\label{twodpdfsection}

	Figure \ref{twodpdfplot} exhibits the joint pdfs for the two
sheet models ($\Btot$ perpendicular and parallel to the sheets---the
``perpendicular'' and ``parallel'' models). The median observed column
density ${\Nobs}_{1/2}$ is twice the assumed $\Nperp$ and the median
observed magnetic ${\Bpar}_{1/2}$ is half the assumed $\Btot$; these
univariate medians are indicated by squares on the two  panels. The
significance of these squares is that half the observed $\Bpar$, and
half the observed $\Nobs$, are smaller and half larger. Finally, the
dashed line in the bottom panel exhibits the median ${\Bpar}_{1/2}$
versus $\Nobs$; we calculate this by extracting the conditional pdf
$\psi(\Bpar | \Nobs)$ versus $\Nobs$ and calculating the median from
its cumulative distribution, thus obtaining $B_{los, 1/2}$ versus
$\Nobs$.

	Figure \ref{twodpdfplot} illustrates a crucial observational
signature at large $\Nobs$ that distinguishes between the two sheet
models. More specifically, for the perpendicular model, large $\Nobs$
goes with small $\Bpar$, and {\it vice-versa} for the parallel model.
For the perpendicular model, {\it all} of the datapoints having $\Nobs$
above its univariate median (indicated by the square) have $\Bpar$ {\it
below} its univariate median. In contrast, for the parallel model {\it most}
($66\%$) of the datapoints have $\Bpar$ {\it above} its univariate median:
as $\Nobs$ gets large, the conditional pdf $\psi( \Bpar \, | \, \Nobs)
\rightarrow {{\Nperp}_0 \over \pi \Nobs^2} ({\Btot}_{,0}^2 -
{\Bpar}^2)^{-1/2}$, which produces the median ${\Bpar}_{1/2} \rightarrow
0.71 \Btot$.

\subsection{Conversion of the intrinsic to observed bivariate
distributions involving turbulent velocity } \label{bivarV}

	In this section, we derive and display the observed bivariate
distributions involving $\psi(\Bpar,\Vntobs)$ and $\psi(\Nobs,\Vntobs)$,
under the assumption that the turbulent velocity is perpendicular to the
magnetic field.
 
\subsubsection {Conversion of the intrinsic \boldmath{$\phi(\Btot,
V_{turb})$} to the observed \boldmath{$\psi( B_{los}, V_{turb,los})$}}
\label{bivarBV}

	The joint pdf $\psi(\Bpar, \Vntobs)$ is the same for both the
parallel and perpendicular models, and indeed does not depend on whether
there are sheets or not, because $\Va$ is perpendicular to $\Btot$ so that
there is a unique relationship ${\Bpar \over \Btot} = \left[1
- \left( V_{turb,los} \over V_{turb} \right) ^2 \right] ^{1/2}$. As
before, we begin with trivariate distributions involving  $\theta$,
$\Btot$ and $\Va$. We eliminate $\theta$ to obtain

\begin{equation} 
\psi(\Bpar, \Vntobs) = \int_{\Vntobs}^{\infty}
{\Vntobs \over \Va^2} \left[1 - {\Vntobs^2 \over \Va^2} \right]^{-1/2}
\phi_{\Btot}\left(\Bpar \left[1 - {\Vntobs^2 \over \Va^2} \right ]^{-1/2}
\right) 
\phi_{\Va}(\Va) d\Va
\end{equation}

	The top panel of Figure \ref{dualvanhb} displays this joint pdf
for delta-function distributions of $B_{tot}$ and $V_{turb}$. Because we
assume that the $B_{tot}$ and $V_{turb}$ are perpendicular, strong
fields go with small velocities and {\it vice-versa}. There are no
contours; they all collapse into a line because, for the delta-function
distributions, there is a one-to-one relationship between $B_{los}$ and
$V_{turb, los}$.

\subsubsection {Conversion of the intrinsic \boldmath{$\phi(\Nperp,
\Va)$} to the observed \boldmath{$\psi( \Nobs, \Vntobs)$} for the
perpendicular model} \label{bivarNVperp}

	In contrast to the above case, the relation between
$\phi(\Nperp, \Va)$ and $\psi( \Nobs, \Vntobs)$ does depend
on the model. For the model with $B_{tot}$ perpendicular to the sheet we
have $\Nobs = {\Nperp \over \cos \theta}$ and $\Vntobs = \Va \sin
\theta$. We obtain 

\begin{equation} \label{nvperp}
\psi(\Nobs, \Vntobs) = \int_{\Nobs}^\infty
{\Nperp \over \Nobs^2} \left[ 1 - {\Nperp^2 \over
\Nobs^2} \right]^{-1/2} 
\phi_{\Va} \left( \Vntobs \left[ 1 - {\Nperp^2 \over
\Nobs^2} \right]^{-1/2} \right) \phi_{\Nperp}(\Nperp) d\Nperp
\end{equation}

	The middle panel of Figure \ref{dualvanhb} displays this joint
pdf for delta-function distributions of $N_{\perp}$ and $V_{turb}$. 
There are no contours; they all collapse into a line because, for the
delta-function distributions, there is a one-to-one relationship between
$B_{los}$ and $V_{turb, los}$. Because we assume that the $B_{tot}$ and
$V_{turb}$ are perpendicular, high HI columns go with small velocities
and {\it vice-versa}. 

\subsubsection {Conversion of the intrinsic \boldmath{$\phi(\Nperp,
\Va)$} to the observed \boldmath{$\phi( \Nobs, \Vntobs)$} for the
parallel model} \label{bivarNVpar}

	The parallel model is more complicated, as it was in \S
\ref{Btotpar}, because $\Phi$ enters explicitly: because $\Va$ is
assumed to be perpendicular to $\Btot$, $\Vntobs = \Va [1 - (\sin \theta
\sin \Phi)^2] ^{1/2}$. The final expression equivalent to \ref{nvperp}
is quite complicated, so we deal with it numerically. The bottom panel of
Figure \ref{dualvanhb} displays this joint pdf for delta-function
distributions of $N_{\perp}$ and $V_{turb}$; the contours come from a
Monte-Carlo calculation.

\subsubsection{Summary} \label{bivardisc}

\begin{figure}[p!]
\begin{center}
\includegraphics[width=5.0in] {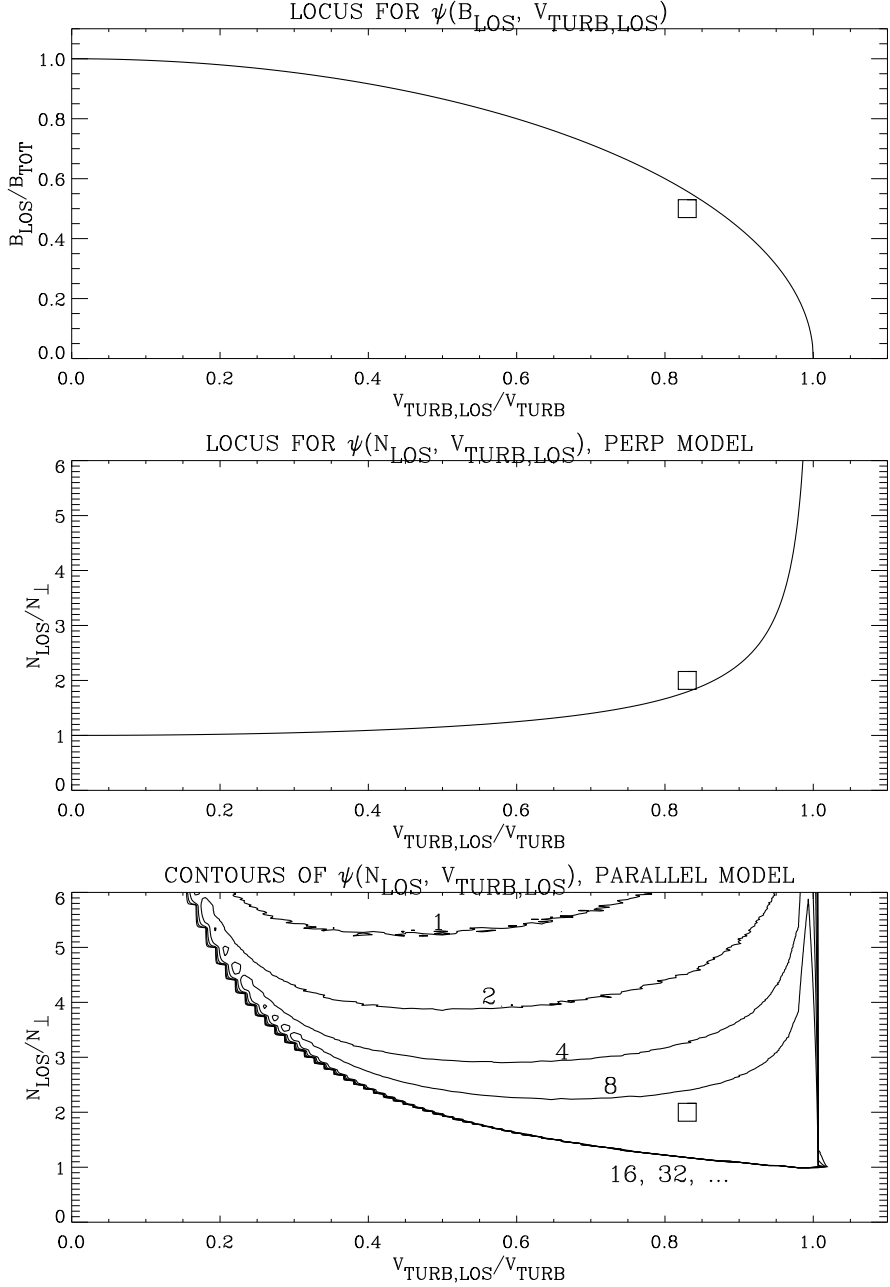}
\end{center}

\caption{Bivariate distributions for the model having turbulent
velocities perpendicular to the magnetic field, discussed in \S
\ref{bivarV}. These distributions assume $\delta$-function distributions
of the astronomical parameters $\Va$, $\Btot$, and $\Nperp$. Top panel,
the locus of contours of the bivariate distribution $\psi(B_{los},
V_{turb,los})$ (observed quantities).  Middle and bottom panels:
contours of $\psi(\Nobs, \Vntobs)$ Squares show the univariate medians.
\label{dualvanhb} } \end{figure}

	The complicated nature of the the univariate distribution 
$\psi(\Vntobs)$ in equation \ref{volterra1} means that the bivariate
distributions that involve $\Vntobs$ are even more complicated and
preclude closed-form solutions when we use the observationally derived
pdfs. However, we can easily present the bivariate results for
delta-function distributions of $\Btot$, $\Va$, and $\Nperp$, shown in
Figure \ref {dualvanhb}. Two, $\psi(\Bpar, \Vntobs)$ and $\psi(\Nobs,
\Vntobs)$, degenerate into lines because the observed variables depend
only on $\theta$.

\section{ DERIVATION OF INTRINSIC ASTRONOMICAL UNIVARIATE PDFs FROM
OBSERVED HISTOGRAMS}

\label{unihist}

\subsection{Derivation of the intrinsic \boldmath{$\phi(B_{tot})$ from
the histogram of observed $B_{obs}$}} 

\label{Btotderivation}

\begin{figure}[p!]
\begin{center}
\includegraphics[width=5.0in] {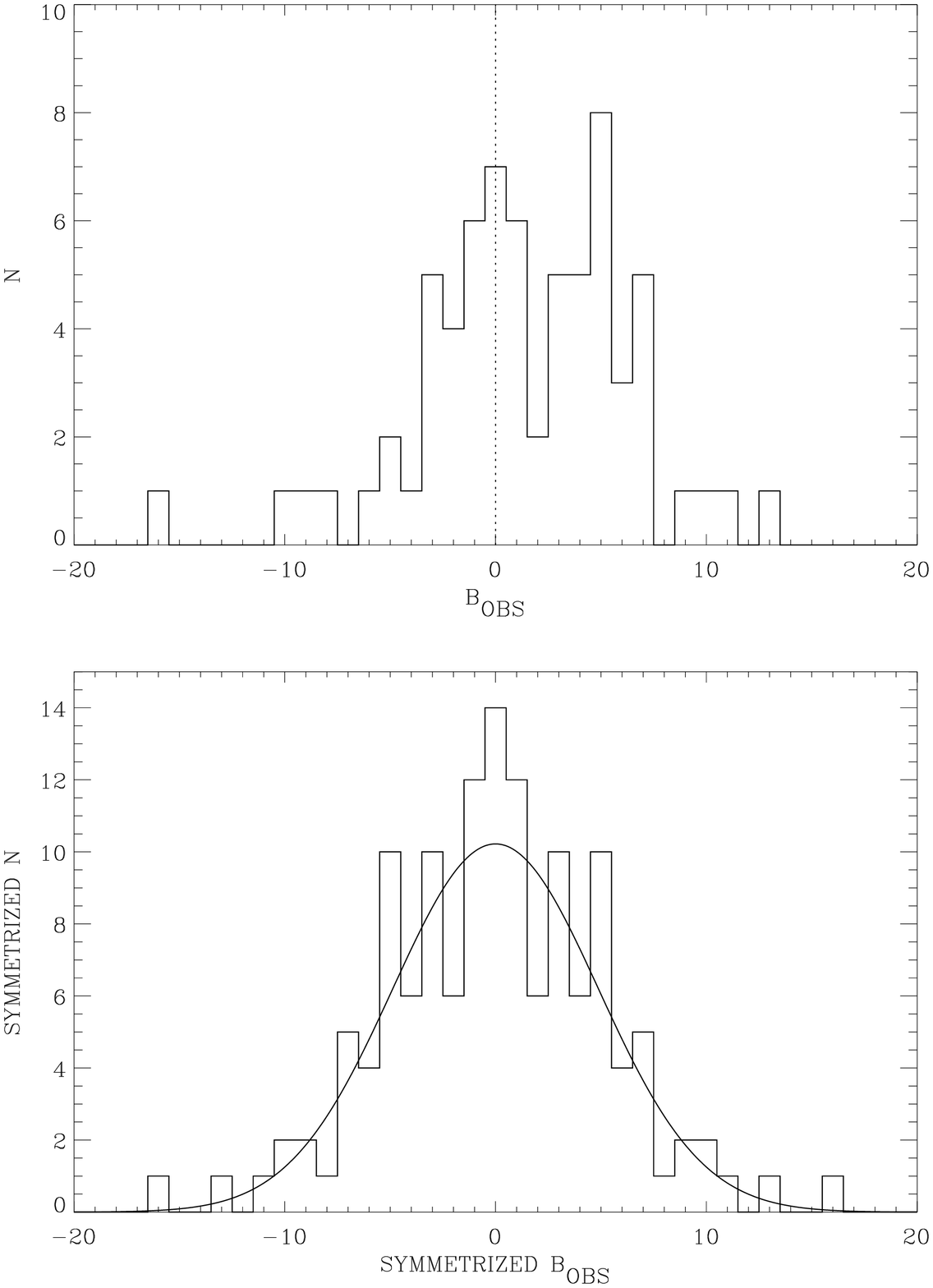}
\end{center}

\caption{Top panel: the histogram of observed field strengths $B_{obs}$
(in $\mu$G). Bottom panel: the same data, but symmetrized with each
observation represented a second time with the opposite sign. The smooth
curve is the Gaussian fit of equations \ref{Bpsiobseqns}. \label{bstat1}
} \end{figure}

	Figure \ref{bstat1}, top panel, exhibits the histogram of
measured field strengths $B_{obs}$. It contains 69 measurements, of
which only 12 have the measurement error $\delta B_{los, m} < 2.5 B_{los,
m}$. This histogram is not symmetric: it has 42 instances of $B_{obs}
> 0$ and 27 instances of $B_{obs} < 0$. For a randomly distributed
angle between line-of-sight and the field direction, one expects the
numbers to be equal; the probability that we obtain this imbalance, or
worse, is given by integrating the binomial (coin-tossing) pdf and is
equal to 0.042. This is fairly low, but is hardly low enough to rule out
the random distribution. However, the small probability might indicate
that the selection of sources in the Arecibo sky does, in fact, involve
non-randomness.

	We proceed on the assumption that the distribution of angles is,
indeed, random. This allows us to compare the histogram of the absolute
value of measured field $|B_{los, m}|$ with the theoretical expectation
for various assumed intrinsic pdfs of $B_{tot}$. Figure \ref{bstat1},
bottom panel, exhibits this distribution, but symmetrized so that every
entry for a positive $B_{obs}$ is matched by one with negative 
$B_{obs}$. The solid line is best fit the Gaussian pdf 

\begin{mathletters} \label{Bpsiobseqns}
\begin{equation} \label{Bpsiobs}
\psi(B_{obs}) = {1 \over {\sqrt{2 \pi B_{obs,0}^2}}} \ 
e^{- (B_{obs}^2 / 2 B_{obs,0}^2)} \ ,
\end{equation}

\noindent with 

\begin{equation} \label{Bpsiobs_5.2}
B_{obs,0} = 5.2 \pm 1.3 \ \mu{\rm G} \ . 
\end{equation}
\end{mathletters}

	To find this best fit and error for $B_{obs,0}$, we numerically
sampled a range of trial $B_{obs,0}$ values, and for each trial value
compared the calculated cumulative distribution with that of the data by
performing the Kolmogorov-Smirnov test. This provides sets of the K-S
statistic $D$ and its associated probability $P_{KS}(B_{obs,0})$ that
the assumed distribution matches the observed one (see Press et al
1997). We determined the best fit for $B_{obs,0}$ by choosing the one
that maximizes $P_{KS}$, which is $P_{KS} = 1.0$ (meaning that a
Gaussian is an excellent fit), and we defined its uncertainty as being
where the $P_{KS}$ falls to $32\%$ of its peak value (thus mimicking the
definition of the $1\sigma$ error for Gaussian statistics). 

	Of course, the $5.2 \pm 1.3$ $\mu$G of equation
\ref{Bpsiobs_5.2} represents the pdf of $B_{obs}$. This is not the same
as the pdf of $B_{los}$ because of observational noise, which is very
significant. Therefore, the dispersion of $B_{los}$ is considerably
smaller. Of course, it is $B_{los}$, not $B_{obs}$, which is the
quantity of interest, so we need to statistically account for the
observational noise. This is not straightforward  because the 69
measurement errors are all different. If they had all been identical,
then we could have used Gaussian statistics and convolutions to derive
the true dispersion ${\Bobs}_{,0}$ from the measured $B_{obs,0}$. 
Because the errors are not identical, we instead use a Monte Carlo
analysis and employ the actual measurement uncertainties instead of
their rms.

\subsubsection{ Monte Carlo Method} \label{montecarlo}

	Here we use the observed values of $B_{obs}$, together with
the uncertainties $\delta B_{noise}$, in a Monte Carlo simulation. We
begin with four different possibilities for the functional form of
$\phi(B_{tot})$; each of these is characterized by the median total
field strength $B_{tot}$, and we calculate 46 uniformly spaced
field possibilities ranging from 1 to 10 $\mu$G. 

	We assume random orientation with respect to the line of sight.
We assume the observational errors $\delta B_{noise}$ to be
Gaussian-distributed. For each test possibility, we perform the set of
69 observations many times (20000 ``trial runs''). Each time, for each
of the 69 measurements we randomly generate a value of $B_{tot}$
according to the assumed $\phi(B_{tot})$; orient it randomly with
respect to the line of sight; derive its observed value $B_{los}$;
generate an uncertainty $\delta B_{los}$ from a Gaussian pdf whose
dispersion is equal to the associated value of $\delta B_{noise}$; and
record the resulting value of the observed $B_{obs} = B_{los} + \delta
B_{los}$, which includes measurement errors. For each trial run, all
four possibilities for the functional form use the same values of
$B_{tot}$, orientation, and $\delta B_{los}$. We make histograms of the
absolute values $B_{obs}$ and use the KS test to compare with the
cumulative histogram of our data. 

\begin{figure}[h!]
\begin{center}
\includegraphics[width=5.0in] {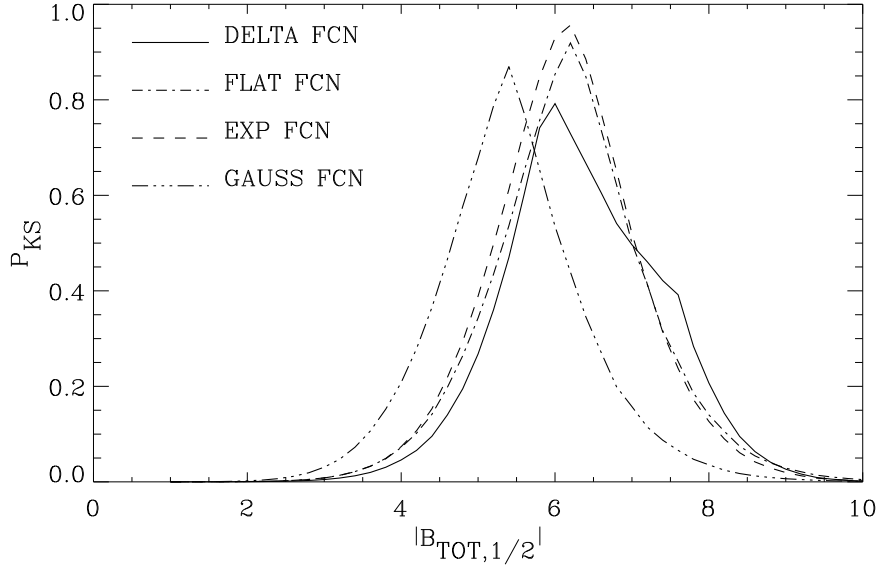}
\end{center}

\caption{ The KS fits of the Monte Carlo simulations for the four trial
functional forms to the data, with the median $|B_{tot,1/2}|$ (in
$\mu$G) as the independent variable. \label{mediantst_abs} }
\end{figure}

	Figure \ref{mediantst_abs} displays the KS fits to the four
functional forms. As anticipated in \S \ref{univar_B}, all four KS fits
are reasonably good, showing that the data cannot distinguish among
these functional forms. However, medians for all four functional forms
are quite similar, so the data do select the median with a reasonably
small uncertainty. 

	To collapse the results of four functional fits into a single
number for the median total field, we simply average the fields for the
four peaks to obtain $6.0$ $\mu$G. To obtain the uncertainty, we use the
maximum spread of the $32\%$ points, which are $4.2$ and $7.8$ $\mu$G.
These combine to make our derived total median field strength

\begin{equation}
|B_{tot, 1/2}| = 6.0 \pm 1.8 \ \mu {\rm G} \ .
\end{equation}

\subsubsection{A Single Uncertainty \boldmath{$\delta B_{noise}$} for
Purposes of Convolution} \label{convolution_adopt}

	Just above, in \S \ref{montecarlo}, we derived the median value
$B_{tot, 1/2} = 6.0$ $\mu$G. Under the assumption that the distribution
of observed values {\it without} noise is Gaussian, then the pdf of
$B_{tot}$ is the weighted exponential of equation \ref{Btotexp}. In this
equation, the median $B_{tot, 1/2} = 6.0$ $\mu$G corresponds to $B_0 =
3.9$ $\mu$G.  

	This value of $B_0= 3.9$ $\mu$G is really $B_{los,0}$, i.e.\ it
refers to what we would see {\it without} noise; in contrast, {\it with}
noise we found $B_{0,obs}=5.2$ $\mu$G. Thus, for this particular pdf for
$B_{tot}$, our ensemble of $B_{obs}$ is equivalent to convolving the
ensemble of actual values $B_{los}$ with a {\it single} Gaussian of
dispersion 

\begin{equation}
\delta B_{noise, sngl} = (5.2^2 - 3.9^2)^{1/2} = 3.4 \ \mu{\rm G} \ .
\end{equation}

	Below, in our interpretation of bivariate histograms of our
results, we will convolve the Monte-Carlo-derived values of $B_{los}$
(which have no noise) with a Gaussian of this dispersion to obtain
simulations of the measured histograms involving $B_{obs}$ (which
includes noise). This is a quick, approximate way to illustrate how
measurement errors affect the histograms of observed values. 

\subsection{Derivation of the intrinsic \boldmath{$\phi(\Nperp)$ from
the histogram of observed $\Nobs$}} \label{nperpderivation}

	We first distinguish between two distinct sets of data for
$N_{los}$ and $V_{los}$. One is the entire sample of CNM components from
Paper II. The second is the sample we have been discussing,
namely that comprising the 69 statistically interesting measurements
of $B_{obs}$; these are drawn from Paper III, restricted to those
having small uncertainties $\delta B_{noise}$. This smaller sample has a
bias eliminating small column densities, which are irrevocably associated 
with large uncertainties in $\delta B_{noise}$.
	
	We should not use the smaller dataset of 69 samples for
discussing the statistics of column density and velocity, because it is
biased. Instead, we use the larger dataset, but with some restrictions.
We include all components from Paper II that satisfy the restrictions
that $|b| > 10^\circ$, $\Vntobs^2 > 0$, and $\Vntobs < 5.25$ km
s$^{-1}$. The first restriction helps to eliminate blending of CNM
components; the second eliminates 5 components for which the errors
happen to yield $T_{kmax} < T_k$; and the third eliminates 3 outliers to
the analytic approximation below. All this leaves us with 138 for the
sample from Paper II. In the ensuing discussion, we will specify which
dataset we are using.

\subsubsection{A Preliminary: latitude dependence of \boldmath{$\Nperp$}}
\label{nhvsbeesection}

	In \S \ref{randomlyoriented} below we will assume that the
sheets are oriented randomly and derive the intrinsic $\phi(\Nperp)$
from the histogram of observed $\Nobs$. Before doing this, however, we
consider whether the distribution is, in fact, random. In particular, we
might expect the sheets to lie parallel to the Galactic plane, in which
case we would expect $\Nobs \propto {1 \over \sin|b|}$, so we discuss
this possibility first. 

\begin{figure}[h!]
\begin{center}
\includegraphics[width=5.0in] {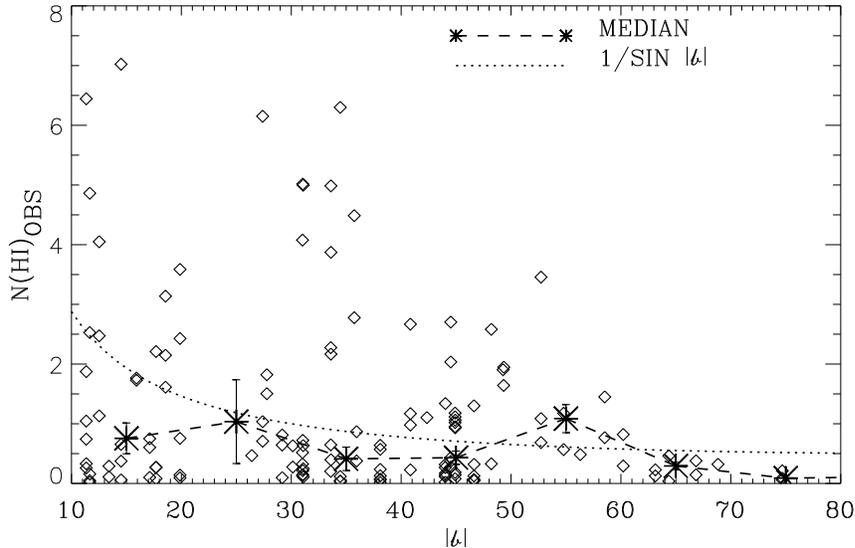}
\end{center}

\caption{Diamonds are datapoints, $\Nobs$  (in $10^{20}$ cm$^{-2}$)
versus $|b|$. Each star represents the median of $\Nobs$ within the
$10^\circ$ wide bin centered on the star; the stars are connected by the
dashed line. The dotted line is $\Nobs = {1. \over \sin|b|}$, and is
meant only to guide the eye. Two points, located at $(|b|, N(HI) =
(28,12)$ and $(19,13)$, are excluded from the figure (but not the
medians) to save space. \label{nh_vs_beefig} } \end{figure}

	Figure \ref{nh_vs_beefig} exhibits the latitude dependence of
$\Nobs$. Looking at the points, one sees some higher values near smaller
$|b|$, leading one to suspect that this might be a statistical trend. To
test this suspicion we examine the latitude dependence of binned
medians. The errorbars on the starred medians were calculated using the
absolute values of the residuals, i.e.\ the error is ${\Sigma |\Nobs -
{\Nobs}_{1/2}| \over M^{1/2} (M-1)}$, where ${\Nobs}_{1/2}$ is the
median and $M$ is the number of points in the bin. The dashed line
connects the medians of $\Nobs$ for $10^\circ$-wide bins in $|b|$. 

	If $\Nobs \propto {1 \over \sin|b|}$, then these medians should
follow the same dependence; we provide the dotted line, which is $\Nobs
= {1 \over \sin|b|}$, for a visual comparison. Visually, there is
absolutely no tendency for the medians to follow $1. \over \sin|b|$. The
presence of datapoints with high $\Nobs$ at small $|b|$ might be real or
it might be a statistical artifact. If it is real, it might mean that
sheets are in fact aligned with the Galactic plane to some degree and/or
have their own intrinsic distribution of $\Nperp$. 

	There is more to this relationship than random statistics,
namely our nonuniform and nonrandom sampling of the sky. Of course, we
are restricted to Arecibo's sky, which is a $38^\circ$-wide swath in
declination. In the vicinity of right ascension $04^h30^m$ it passes
through the Taurus/Perseus region at $\ell \sim 180^\circ$, covering a
large spread in latitude and a small range in longitude. Large column
densities persist here up to $|b| \sim 50^\circ$, which is an anomaly in
the overall latitude distribution of the neutral ISM. Thus in
Taurus/Perseus, if we were to plot the {\it total} observed column
density towards any direction (as opposed to $\Nobs$, which is the
observed column density in one CNM Gaussian component) versus $|b|$, it
would not behave anything like $1. \over \sin|b|$. On the contrary, in
other longitude ranges the situation differs. For example, in the
vicinity of $\ell= 240^\circ$ the ISM column density is anomalously
small. Thus, plotting quantities versus $|b|$ also plots them versus a
biased distribution in $\ell$. Accordingly, given the nonuniform
sampling of the sky, it is very difficult to establish a latitude
dependence.

	From the above, we conclude that we cannot extract a latitude
dependence of $\Nobs$ from the data. Accordingly, we proceed under the
simplest assumption, namely that the sheets are randomly oriented with
respect to the observer's line of sight.

\subsubsection{Derivation of \boldmath{$\phi(\Nperp)$} assuming the
sheets are randomly oriented} \label{randomlyoriented}

\begin{figure}[p!]
\begin{center}
\includegraphics[width=4.5in] {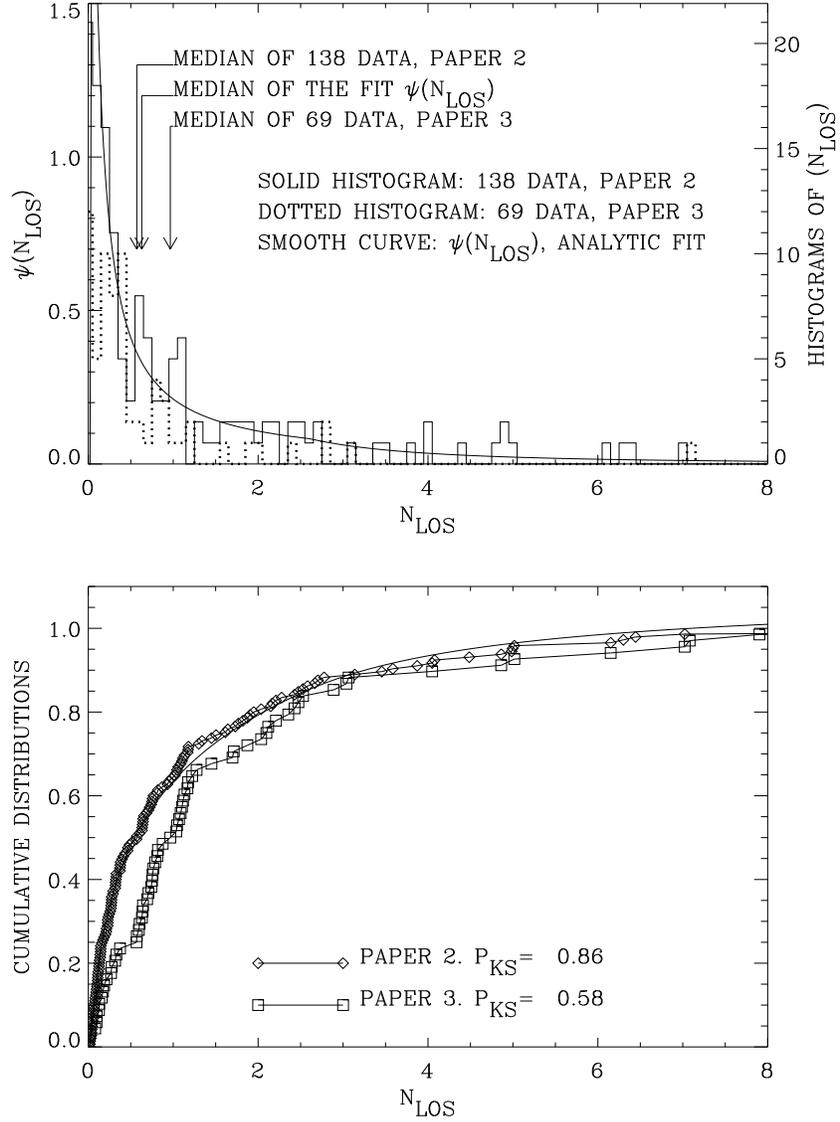}
\end{center}

\caption{Analytic fit and observed distributions for $\Nobs$ (in
$10^{20}$ cm$^{-2}$). The solid histogram in the top panel is for the
full set of 138 sheets from Paper II; the dashed is for the restricted
set of 69 from Paper III, used in the present paper. The smooth curve is
$\psi( N_{los})$, the pdf of equation \ref{Nobsobs}, which is the
analytic fit to the 138 datapoints; this curve closely follows $\Nobs
^{-1}$. The bottom panel shows the three equivalent cumulative
distributions; the diamonds are the solid histogram and the squares the
dotted one. \label{nhstat} } \end{figure}

	The top panel of Figure \ref{nhstat} exhibits the histograms of
observed column densities $\Nobs$ together with a curve that, over most
of the region, goes roughly as $\Nobs ^{-1}$  (see equation
\ref{Nobsobs} below). The solid histogram is for the full Paper II
sample of 138 datapoints.  The eyeball tells us that this is a
reasonably good fit. Using equation \ref{Npsiphi}, we find that if
$\psi(\Nobs) \propto \Nobs ^{-1}$, then $\phi(\Nperp)$ has the same
dependence, namely $\phi(\Nperp) \propto \Nperp ^{-1}$. Unfortunately,
its integral diverges so it is not a valid pdf. Rather, we must impose
lower and upper limits on this pdf.

	To proceed, we assume

\begin{eqnarray} \label{Nperp}
\phi[\Nperp]= 
\left\{ 
\begin{array}{ll}
\vspace{4pt}
{\kappa \over \Nperp } & 
	{\rm if} \  {\Nperp}_{min} \leq \Nperp \leq {\Nperp}_{max} \\ 
        0       & {\rm otherwise}
\end{array}
\right. 
\end{eqnarray}

\noindent where $\kappa = [\ln ( {\Nperp}_{max} / {\Nperp}_{min})]^{-1}$
and derive the corresponding $\psi(\Nobs)$ from equation \ref{Npsiphi}.
This yields

\begin{eqnarray} \label{Nobsobs}
\psi(\Nobs)= 
\left\{ 
\begin{array}{ll}
\vspace{4pt}
0  & \Nobs < {\Nperp} _{min} \\
\vspace{4pt}
\kappa {\Nobs - {\Nperp}_{min} \over \Nobs ^2} & 
{\Nperp}_{min} \le \Nobs \le {\Nperp}_{max} \\
\kappa {{\Nperp}_{max} - {\Nperp}_{min} \over 
\Nobs ^2} &  \Nobs > {\Nperp}_{max} 
\end{array}
\right. 
\end{eqnarray}

\noindent We then numerically cover a grid of trial values for
$({\Nperp}_{min}, {\Nperp}_{max})$ and for each combination derive the
corresponding $\psi(\Nobs)$. From that, we calculate the cumulative
distribution and compare with the observed one by performing the K-S
test, from which we obtain the bivariate probability
$P_{KS}({\Nperp}_{min}, {\Nperp}_{max})$ that the assumed distribution
matches the observed one. We determine the best fit parameters by
choosing the combination $({\Nperp}_{min}, {\Nperp}_{max})$ that
maximizes $P_{KS}$.

\begin{figure}[h!]
\begin{center}
\includegraphics[width=3.0in] {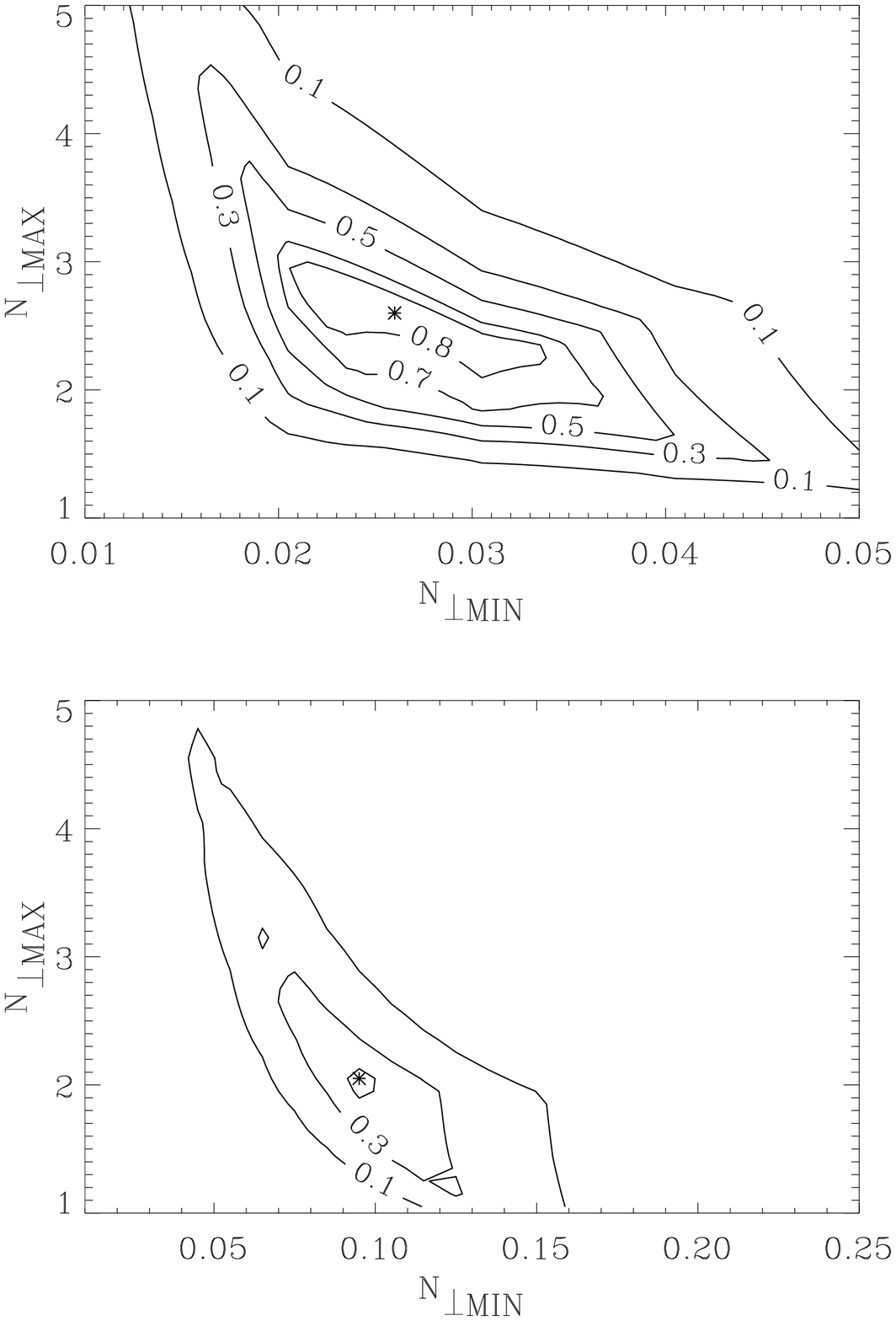}
\end{center}

\caption{Contours of $P_{KS}({\Nperp}_{min}, {\Nperp}_{max})$ (units are
$10^{20}$ cm$^{-2}$). The stars are the adopted solutions for
$({\Nperp}_{min}, {\Nperp}_{max})$. Top panel, the 138 datapoints from
Paper II; bottom panel, the 69 datapoints from paper III.
\label{plotbruteks} } \end{figure}

	Figure \ref{plotbruteks} shows contour plots of
$P_{KS}({\Nperp}_{min}, {\Nperp}_{max})$. The top panel is for the 138
datapoints from Paper II, which is the larger and unbiased sample. The
highest contour encloses a well-defined area that has two peaks with
$P_{KS} \sim 0.89$ connected by a saddle. We don't regard the
differences within the highest contour to be significant and we adopt the
point indicated by a star as the best solution; here $P_{KS} = 0.86$. We
adopt its uncertainty as defined by the contours where $P_{KS}$ drops to
$\sim 0.30$, which is $\sim 32\%$ of the peak value 0.89. This yields

\begin{mathletters} \label{Nobsobsnumbers}
\begin{equation}
{\Nperp}_{min} = 0.026 _{- 0.010}^{+0.019} \times 10^{20} \ {\rm cm}^{-2}
\end{equation}
\begin{equation}
{\Nperp}_{max} = 2.6  _{- 1.2}^{+1.9} \times 10^{20} \ {\rm cm}^{-2}
\end{equation}
\end{mathletters}

	These two limits differ by two orders of magnitude! For most of
the range of $\Nperp$ between these limits, $\phi(\Nperp) \propto \Nperp
^{-1}$ to a very good approximation. In Figure \ref{nhstat}, the smooth
solid line in the top panel is equation \ref{Nobsobs} with these values
of $({\Nperp}_{min}, {\Nperp}_{max})$. The bottom panel shows the
cumulative distributions, both for the data (in diamonds) and for this
derived pdf. To the eye the match looks excellent, and this is confirmed
by the high value of the K-S probability $P_{KS}=0.86$.

	Figure \ref{nhstat} also shows the distributions for the current
magnetically-selected sample 69 datapoints from Paper III. The top panel
shows the histogram in a dotted line. The dotted histogram differs
significantly from the solid one because it doesn't have the large
increase for small $\Nobs$. The magnetic selection excludes datapoints
whose uncertainty $\delta \Bpar$ exceeds an upper limit. This
restriction biases the column densities, because it is impossible to
obtain small uncertainties on $\Bpar$ for small $\Nobs$. Consequently,
the pdf of $\Nobs$ is cut off at small values. The bottom panel of
Figure \ref{plotbruteks} shows contour plots of $P_{KS}({\Nperp}_{min},
{\Nperp}_{max})$ for this set of 69 datapoints. Clearly, the contours
peak at a different location. The bias in the magnetically selected
points is reflected in the larger value of ${\Nperp}_{min}$, 0.090 for
the magnetically-selected sample versus 0.026 for the unbiased sample.
The values of ${\Nperp}_{max}$ are comparable: 2.0 and 2.6 for the small
and large samples, respectively. The maximum value of $P_{KS}= 0.58$,
indicating good agreement between the data and the analytic
representation.
 
\subsection{Derivation of the intrinsic \boldmath{$\phi(\Va)$ from
the histogram of observed $\Vntobs$}} \label{vaderivation}

\begin{figure}[h!]
\begin{center}
\includegraphics[width=4.5in] {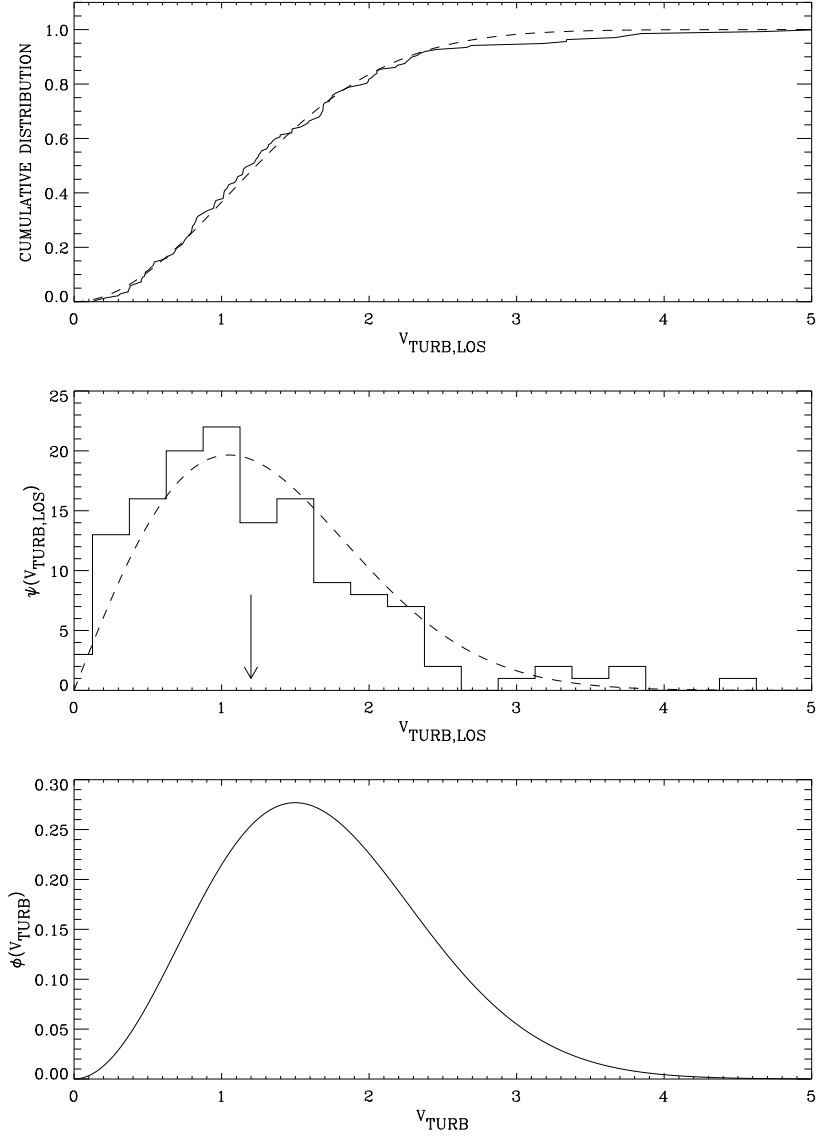}
\end{center}

\caption{The top panel shows the fitted and actual cumulative
distributions of $V_{turb,los}$ (in km s$^{-1}$). The middle panel shows
the corresponding pdfs, i.e.\ the data histogram and
$\psi(V_{turb,los})$. The third panel shows our fitted $\phi({\Va})$.
The arrow shows the median of the observed histograms. Line widths are
dispersions. \label{vnt2e} } \end{figure}

	The top panel of Figure \ref{vnt2e} exhibits the cumulative
distribution of the 138 $V_{turb,los}$ from Paper II (solid line),
together with an adopted cumulative distribution (dashed line). We
obtained the fit by trial and error: for trial functional forms for
$\phi(V_{turb})$, we used equation \ref{volterra1} to calculate the
corresponding $\psi(V_{turb,los})$. Trial functional forms included
$\phi(V_{turb}) \propto x^n e^{-x}$, with $n=(0.25, 0.5, 1., 2.)$, where
$x = {V_{turb} \over V_{turb,0}}$. The value $n=1$ is significantly
better than the others, resulting in $P_{KS} = 0.91$ for $V_{turb,0} =
1.06$ km s$^{-1}$. This functional form is not unique: the function 
$\phi(V_{turb}) \propto x^{1/2} e^{-x^2}$ fits almost as well, with
$P_{KS} = 0.81$ for $V_{turb,0} = 1.7$ km s$^{-1}$. These line widths
are dispersions. The middle panel shows the data histogram and
corresponding pdf $\psi(V_{turb,los})$, and the arrow indicates the
median of the observed distribution at $V_{turb,los,1/2}=1.2$ km
s$^{-1}$. The bottom panel shows $\phi(V_{turb})$.

	The above paragraph assumes the model in which $V_{turb}$ is
perpendicular to $B_{tot}$. If we drop that assumption, then the
line-of-sight distribution of $V_{turb,los}$ has the same functional
form as that of $V_{turb}$ (but is narrower by $3^{1/2}$). A good fit
for the observed distribution is $\psi(V_{turb,los}) \propto x^2 e^{-x}$,
i.e.\ $n=2$ in the above paragraph, with $V_{turb,los,0}=0.45$ km
s$^{-1}$. This line width is dispersion. 

\section{DERIVATION OF PREDICTED BIVARIATE OBSERVED DISTRIBUTIONS AND
COMPARISON WITH DATA} \label{bihist}

	In this section, we use the previous determinations of the pdf
of intrinsic quantities $\phi(B_{tot})$, $\phi({N_\perp})$, and
$\phi(V_{turb})$ to determine the expected bivariate distributions of
observed pairs $\psi(B_{obs}, N_{los})$, $\psi(V_{turb,los}, B_{obs})$,
and $\psi(V_{turb,los}, N_{los})$. If the observations were good enough,
this would enable us to use the correlations among observed quantities
to infer astrophysical model parameters. Unfortunately, we will find
that we cannot.

\subsection{ \boldmath{$\phi(\Btot, \Nperp$)} and
\boldmath{$\psi(B_{obs}, \Nobs$)} } 

\label{bivarBNobs}

	Here we employ the observationally-derived univariate
distributions for $\Btot$ and $\Nperp$ to predict the bivariate
distribution of the observed quantities $\psi(\Bpar, \Nobs)$, for the
two cases of $\Btot$ perpendicular and parallel to the sheets. However,
this single step does not provide a firm basis for comparison with the
data because the observational uncertainties on $\Bpar$ are so large.

As we saw in \S \ref{montecarlo}, the noise precludes our deriving even
the univariate $\phi(B_{tot})$, so for discussion purposes we adopt the
EXP FCN of equation \ref{Btotexp} with $B_0 = 3.9$ $\mu$G (which provides
the required median value 6.0 $\mu$G) (\S \ref{montecarlo}). Even with
this, however, there is additional step: we must predict the bivariate
distribution of the {\it measured} $B_{obs}$ (which includes
observational error), i.e.\ we must obtain $\psi(B_{obs},N_{los})$
instead of $\psi(B_{los}, N_{los})$. We accomplish this by convolving
the conditional distribution $\psi(\Bpar \, | \, \Nobs)$ with a Gaussian
having the rms measurement dispersion 3.4 $\mu$G (\S
\ref{convolution_adopt}), and doing this as a function of $N_{los}$.

\subsubsection{Case of \boldmath{$\Btot$} Perpendicular to the Sheet}
\label{bivarBNobsperp}

	Here we use equation \ref{perpsheet} to predict the observed
$\psi(\Bpar, \Nobs)$ for the perpendicular case.  As explained above,
for $\phi(B_{tot})$ we use the EXP FCN of equation \ref{Btotexp} with
$B_0 = 3.9$ $\mu$G ; and $\phi_{\Nperp}$ comes from \S 
\ref{nperpderivation}. The analytic solution is somewhat cumbersome and
yields

\begin{eqnarray} \label{obsperpcase}
\psi(\Bpar,\Nobs)= 
\left\{ 
\begin{array}{ll}
\vspace{4pt}
{\kappa {\sqrt{2}} \over {\sqrt{\pi}} B_0 \Nobs}
	
	\left[	e^{ -[\Bpar ^2 / (2 B_0^2)]} - 
	e^{ - [ \Bpar^2 \Nobs^2  / ( 2 B_0^2 {\Nperp}_{min}^2)] \ } \right]
& \Nobs \leq {\Nperp}_{max} \\
{\kappa {\sqrt{2}} \over {\sqrt{\pi}} B_0 \Nobs} 
	 \left[
	e^{ - [\Bpar^2  \Nobs^2 /  ( 2 B_0^2 {\Nperp}_{max}^2)] }  - 
	e^{ - [\Bpar^2  \Nobs^2  / ( 2 B_0^2 {\Nperp}_{min}^2)] \ } \right]
& \Nobs > {\Nperp}_{max}
\end{array}
\right. 
\end{eqnarray}

\begin{figure}[h!]
\begin{center}
\includegraphics[width=5.0in] {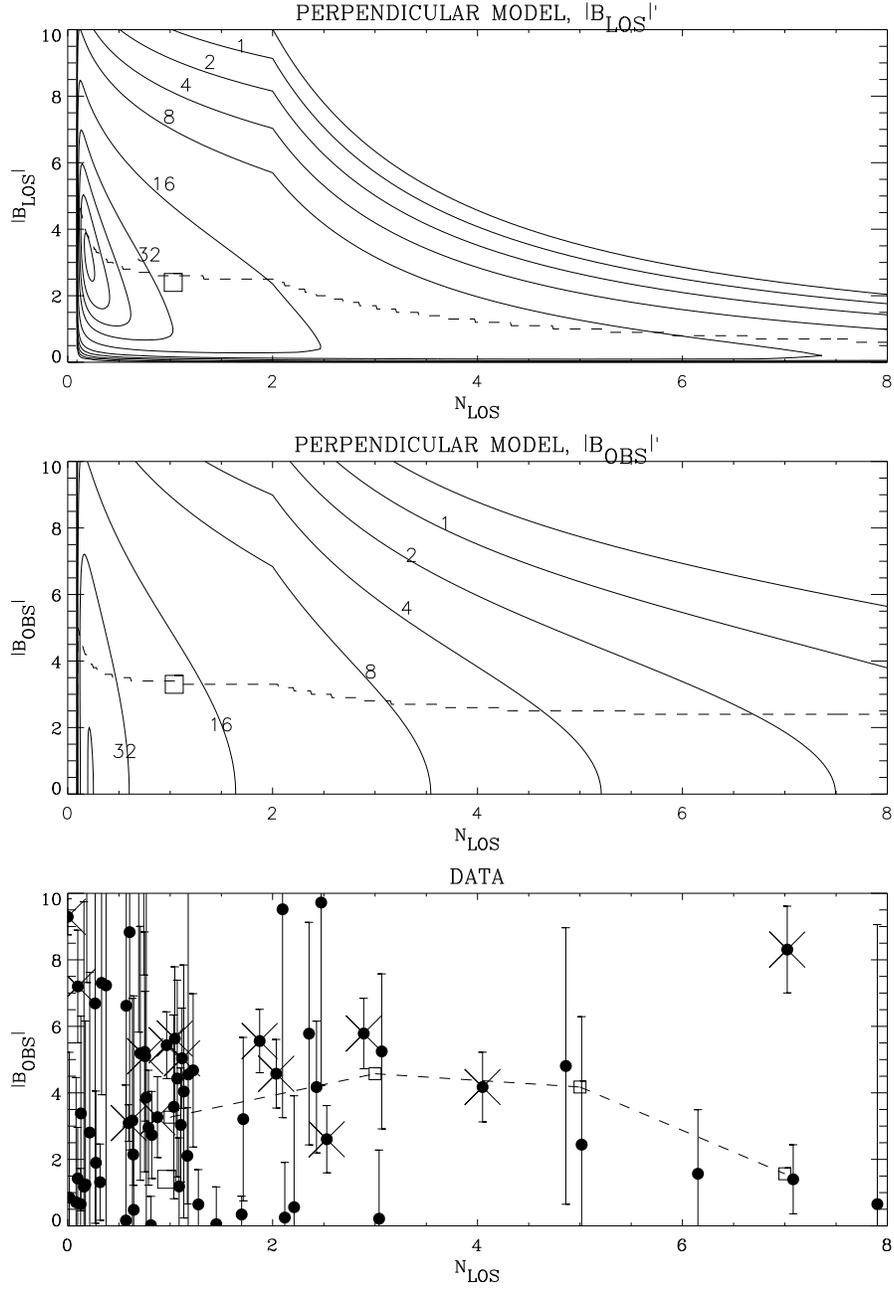}
\end{center}

\caption{Top panel, contours from equation \ref{obsperpcase} of
$\psi(\Bpar,\Nobs)$ for the perpendicular case derived using the
observationally-derived $\phi(\Btot, \Nperp)$.  Units are $\mu$G and    
$10^{20}$ cm$^{-2}$. The middle panel shows $\psi(B_{obs},\Nobs)$,
which includes measurement errors on $\Bpar$.  The square marks the
univariate medians. Bottom panel, the data. \label{cplotperp} } 
\end{figure}

\noindent This produces contours in the $(\Bpar,\Nobs)$ plane; Figure
\ref{cplotperp}, top panel, shows the results. The version including
measurement errors is in the middle panel. 

	The general trend in the top panel is clear from the discussion
of \S \ref{twodpdfplot}: large measured column densities $\Nobs$ produce
small measured fields $\Bpar$ because the field lines are perpendicular
to the sheets. Unfortunately, including the observational uncertainties,
as we do in the middle panel, obscures this trend. We defer further
discussion to \S \ref{bettersheetmodel}.

\subsubsection{Case of \boldmath{$\Btot$} Parallel to the Sheet}
\label{bivarBNobspar}

\begin{figure}[h!]
\begin{center}
\includegraphics[width=5.0in] {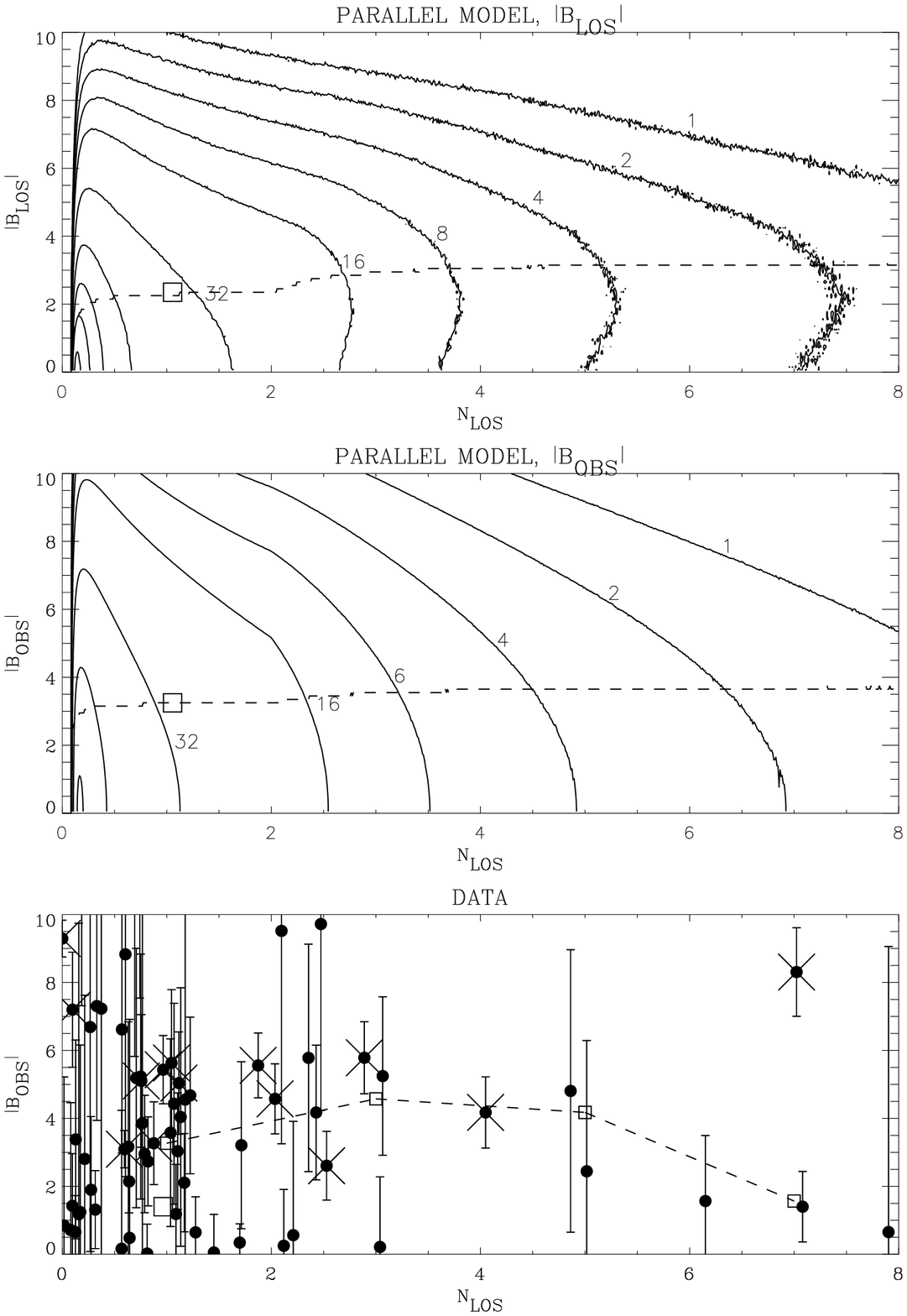}
\end{center}

\caption{Top panel, contours of $\psi(\Bpar,\Nobs)$ from a Monte Carlo
simulation for the parallel case derived using the
observationally-derived $\phi(\Btot, \Nperp)$.  Units are $\mu$G and    
$10^{20}$ cm$^{-2}$. The middle panel shows $\psi(B_{obs},\Nobs)$,
which includes measurement errors on $\Bpar$. The square marks the
univariate medians. Bottom panel, the 69 datapoints from Paper III.
\label{plotdualpar} }  \end{figure}

	Here we proceed as in \S \ref{bivarBNobsperp}, but use equation
\ref{parsheet} instead of equation \ref{perpsheet} to predict the
observed $\psi(\Bpar, \Nobs)$ for the parallel case. The math is
complicated, so we proceed by calculating $\psi (\Bpar,\Nobs)$  using a
Monte Carlo simulation. Figure \ref{plotdualpar} shows the results. The
version including measurement errors is in the middle panel.

	The general trend in the top panel is clear from the discussion
of \S \ref{twodpdfplot}: for $\Nobs \gg {\Nperp}_{max}$, we see sheets
more nearly edge-on where $\Bpar$ is usually large and the conditional
pdf $\psi(\Bpar \, | \, \Nobs)$ becomes independent of $\Nobs$. We defer
further discussion to \S \ref{bettersheetmodel}.

\subsection{ \boldmath{$\phi(\Va, \Btot$)} and \boldmath{$\psi(\Vntobs,
\Bpar$)} }

\label{bivarVBobs}

\begin{figure}[p!]
\begin{center}
\includegraphics[width=5.0in] {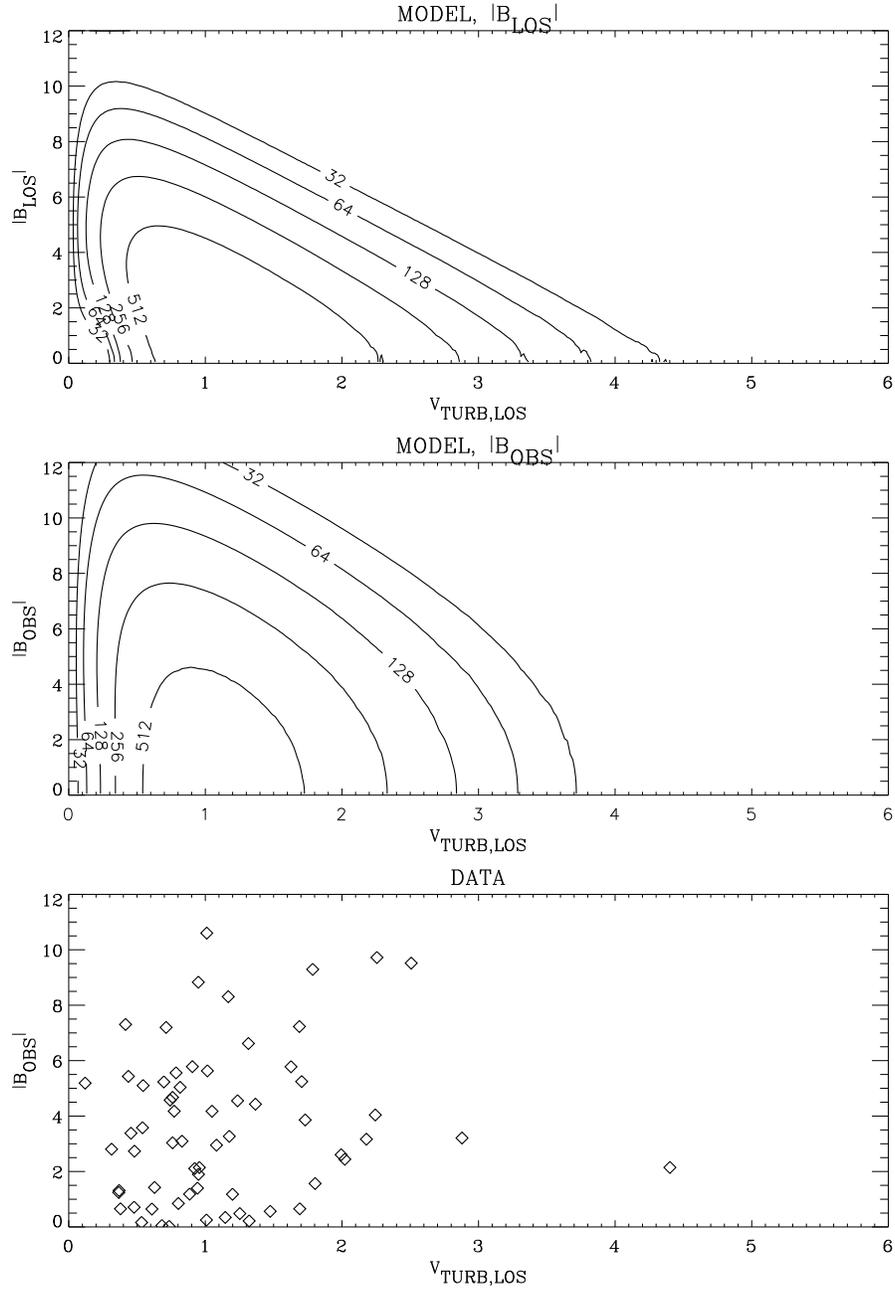}
\end{center}

\caption{Top panel, contours of $\psi(\Vntobs,\Bpar)$ from a Monte Carlo
simulation using the observationally-derived $\phi(\Va, \Btot)$. Units
are km s$^{-1}$ and $\mu$G.  The middle panel shows
$\psi(\Vntobs,B_{obs})$, which includes measurement errors on $B_{los}$.
Bottom panel, the data. \label{contourbva} }  \end{figure}

	As noted in \S \ref{bivarBV}, the bivariate distribution does
not depend upon any model regarding the geometrical shape of the HI
clouds.  Figure \ref{contourbva}, top panel, shows contours of
$\psi(\Vntobs,\Bpar)$ from a Monte Carlo simulation using the
observationally-derived $\phi(\Va, \Btot)$; these $B_{los}$ do not
include measurement errors. The middle panel shows
$\psi(\Vntobs,B_{obs})$ and the bottom panel shows the data. Contours
are spaced by a factor of two. The eyeball says that the fit to these 69
datapoints is not bad, which makes it tempting to conclude that the
turbulence is, in fact, perpendicular to the magnetic field.
Unfortunately, however, the eyeball also says that the distribution of
points doesn't differ much from one in which $|B_{obs}|$ and
$V_{turb,los}$ are uncorrelated, meaning that such a conclusion is
unwarranted. 

\subsection{ \boldmath{$\phi(\Va, \Nperp$)} and \boldmath{$\psi(\Vntobs,
\Nobs$)} }

\label{bivarVNobs}

\begin{figure}[p!]
\begin{center}
\includegraphics[width=5.0in] {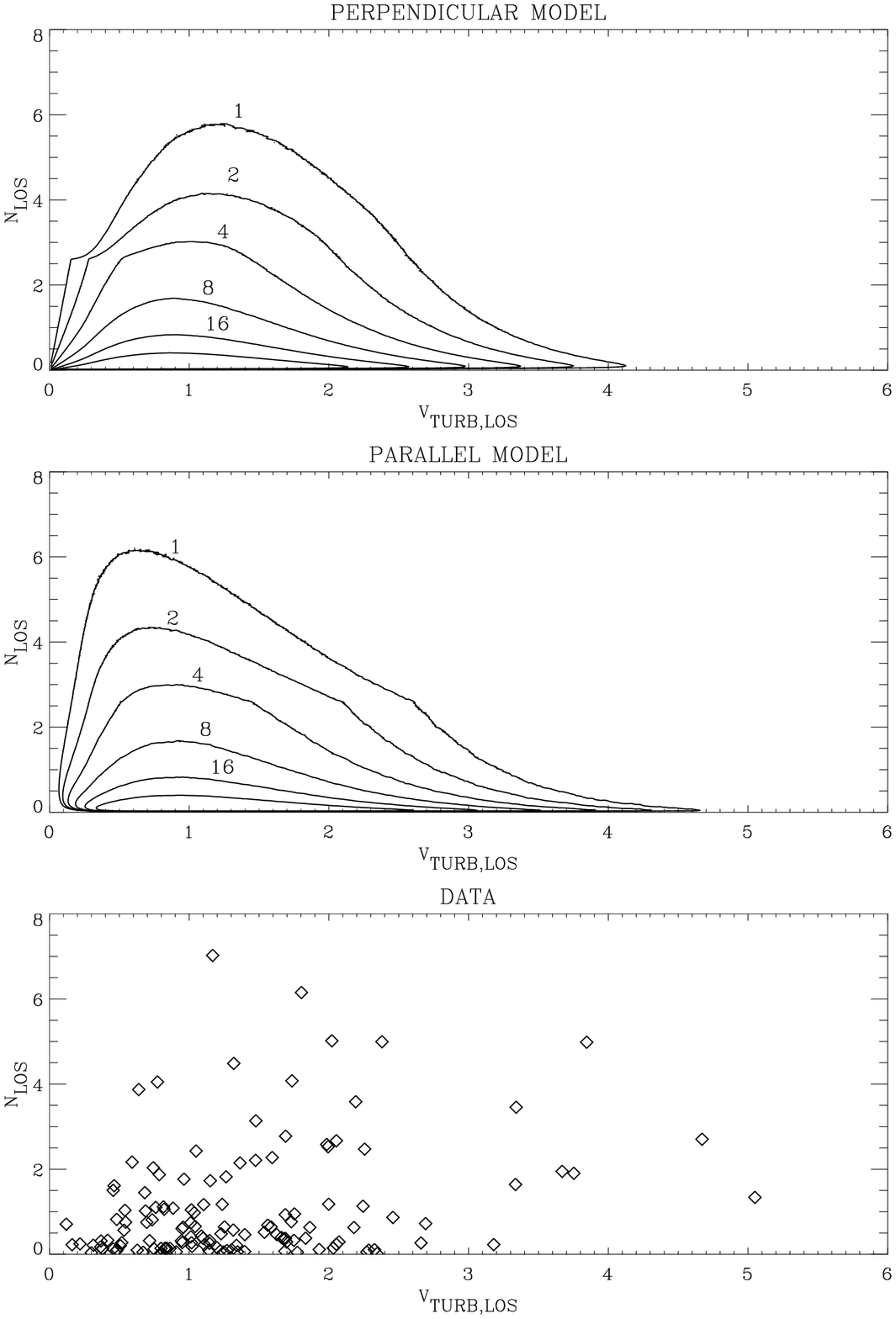}
\end{center}

\caption{Contours of $\psi(\Vntobs,\Nobs)$ from a Monte Carlo simulation
using the observationally-derived $\phi(\Va)$ and $\phi(\Btot)$.   Units
are km s$^{-1}$ and $10^{20}$ cm$^{-2}$. Top and middle panels show the
perpendicular and parallel sheet models, respectively.  Bottom panel
shows the data. \label{contournva} }  \end{figure}

	Here, as above with the relationship between $\Bpar$ and
$\Nobs$, the predicted bivariate distribution depends on whether the
magnetic field is perpendicular or parallel to the sheet. 
Figure \ref{contournva}, top panel, shows contours of
$\psi(\Vntobs,\Bpar)$ for the perpendicular model; the middle panel
shows the parallel model. The contours are derived from Monte Carlo
simulations using the observationally-derived $\phi(\Va, \Nperp)$,
assuming that two quantities are independent. The bottom panel shows the
data. 

	Unfortunately, the difference between the contours in the top
and middle panel is not large and it doesn't take a quantitative
analysis to tell that the data cannot distinguish between the two
models. To the eye, the data look in reasonable agreement with both
models. 

\section{WHICH SHEET MODEL FITS BETTER?} \label{bettersheetmodel}

	The key to distinguishing between the two sheet models is the
bivariate distributions because the univariate distributions don't
depend on whether the perpendicular or parallel model reigns.  Of the 
three bivariate distributions, two are relevant.

	Figure \ref{contournva} displays $\psi(\Vntobs,\Nobs)$.
Unfortunately, the differences between the contours in the top panel
(perpendicular model) and middle panel (parallel model) are not large.
We can make a quantitative comparison of these bivariate distributions
with the data by using the two-dimensional generalization of the K-S
test described by Press et al (1997), which provides $P_{KS}$ in a
similar fashion as does the usual one-dimensional K-S test. The values
are $P_{KS}= 0.032$ and 0.016 for the perpendicular and parallel cases,
respectively. These values are both small, but not so so small as to
rule out agreement of the data with the model. They differ by a factor
of two, but this is nowhere near enough of a difference to distinguish
between the two cases. Moreover, our basic assumption in deriving these 
distributions is the anisotropy of the turbulent velocity $\Va$, which
might not be correct. Unfortunately, this bivariate distribution cannot
distinguish between the two models.

	The bivariate pair $\psi(\Nobs,\Bpar)$  is more directly related
to the two models because there is no  additional assumption involving
anisotropy of the turbulence.  Figures \ref{cplotperp} and
\ref{plotdualpar} exhibit the pair for the perpendicular and parallel
cases, respectively. After including the measurement errors (middle
panels), the two bivariate distributions look similar, as do the runs of
the median $B_{obs,1/2}$ versus $\Nobs$. Neither represents the data
very well. We can be quantitative and perform the two-dimensional K-S
test, as above; the perpendicular and parallel cases yield $P_{KS} =
0.022$ and 0.024, respectively. Again, the values are small but not so
small as to rule out the models; and, again, the test cannot distinguish
between the two cases. 

	One can argue that the single reliably-detected point at
$(\Nobs,\Bpar) \sim (7, 8 \ \mu{\rm G})$ is more consistent with the
parallel model, for which the two quantities tend to be correlated. 
This particular Gaussian component of 3C142.1 (Paper III) has among the
largest values for both quantities, which means that for the
perpendicular model we need either a {\it very} thick sheet and/or a
{\it very} high intrinsic field strength. But we cannot extend this
argument to other sources, whose values are more representative.

	A confident choice of model requires more and better data. 
Unfortunately, this is not likely given current instrumentation, because
we have already used $\sim 800$ hours of Arecibo time for this project. 

\section{ASTROPHYSICAL DISCUSSION: MAGNETIC FIELDS} \label{magdiscussion}

\subsection{Observational issues}

	As discussed in \S \ref{univar_B}, we observe the line-of-sight
component of magnetic field $\Bpar$, which is always less than the
actual magnetic field $\Btot$. Figure \ref{pdf_fig0} and its associated
discussion shows that, in practice, we cannot determine the intrinsic
distribution $\phi(B_{tot})$ from observations of $B_{los}$.
Fortunately, the median is well-determined, with $|B_{tot, 1/2}|= 6.0
\pm 1.8$ (\S \ref{montecarlo}). 

	Figure \ref{pdf_fig0} and Table \ref{medianmean} show that
the median line-of-sight field strength $B_{los,1/2}$ is less than half
that of the total field $B_{tot,1/2}$, with most of the histograms of
$\psi(|B_{los}|)$ peaking at zero.  This functional behavior of
$B_{los}$, which results purely from geometry, is responsible for the
large number of nondetections in our sample. Moreover, it explains
particular cases where $\Bpar$ is small. A spectacular example is the
local-arm field seen against Cas A, $\Bpar = -0.3 \pm 0.6$ $\mu$G. This
surprisingly small result is perfectly consistent with statistical
expectation. Of course, we cannot rule out that the field actually is
really small in any particular case like this, but one needs additional
data to draw such a conclusion! 

	Now consider the large set of magnetic fields observed in 21-cm
line {\it emission} in morphologically obvious structures, reviewed by
Heiles \& Crutcher (2005).  The term ``morphologically obvious'' means
filaments or edge-on sheets.  Edge-on sheets should be edge-on shocks in
which the field is parallel to the sheet.  Here, the statistics reverse
and favor relatively large $\Bpar$.  As explained in \S
\ref{twodpdfsection}, as the line of sight becomes parallel to the
sheet---i.e., for a morphologically obvious sheet---the median
${\Btot}_{1/2} \rightarrow 0.71 \Btot$.  For these structures, measured
fields are strong, ranging from $\sim 5$ to $\sim 10$ $\mu$G.  This is
not inconsistent with a uniform $\Btot \sim 10$ $\mu$G, which is almost
a factor of two above the median CNM field strength.  This suggests that
shocks enhance the field strength, but not by large factors. 

        Finally, compare the CNM median field ${\Btot}_{1/2}$ with other
estimates of field strength. Beck (2003) reviews the most recent
estimate of field strength derived from synchrotron emission, minimum
energy arguments, measured cosmic ray flux, and polarization. He finds
the regular component to be $\sim 4$ $\mu$G and the total component to
be $\sim 6$ $\mu$G. Pulsars give a much smaller value for the regular  
component, but they provide an underestimate if field and electrons are 
uncorrelated, as is likely (Beck et al 2003); nevertheless, they give  
about 5 $\mu$G for the total component (review by Heiles 1996).

        The difference between the regular and total components is the
fluctuating component, whose turbulent spectrum covers a wide range of
scales ranging up to at least tens of parsecs (e.g., the North Polar
Spur).  Our CNM structure sizes are typically of order tenths of a
parsec (Paper III), smaller than much of the magnetic field's turbulence
scale range. For this reason we think that it is more appropriate to 
compare the CNM field strengths with the total component given by Beck,
not the regular one. Our CNM median of $\sim 6.0$ $\mu$G is close to the
local Galactic total component of $\sim 6$ $\mu$G.

\subsection{Astrophysical issues}

	We find that $\Btot$ in the CNM is comparable to the field
strength in other ISM components. This is at first surprising because
the volume density $n(HI)$ in the CNM greatly exceeds that in all other
interstellar structures except molecular clouds.  Flux freezing applies
almost rigorously in the diffuse gas, even in the HI, and as the
interstellar HI changes from CNM to WNM and back again, whether by
thermal instability or dynamical processes, the transition must occur
under the constraints imposed by flux freezing. Under the usual
flux-freezing ideas, magnetic field strength should increase with volume
density. If this increase would actually occur, then we would expect
higher field strengths in the CNM than in other diffuse gas phases
because the ISM should exhibit approximate thermal pressure equality
among the phases. This evidently doesn't happen. In fact,  this absence
of field strength increase for small $n(HI)$ is well known from past
studies  (e.g.\ Crutcher, Heiles, \& Troland 2003, section 3.4), so this
is hardly news.

	The field is strong enough to dominate the gas pressure, and
therefore the dynamics. With a CNM median field $\Btot \sim 6.0$
$\mu$G, which also applies elsewhere in the interstellar volume, the
magnetic pressure is ${P_{mag} \over k} \sim 10400$ cm$^{-3}$ K. This
dominates the CNM pressure $P_{CNM} \sim 3000$ cm$^{-3}$ (Jenkins \&
Tripp, 2001; Wolfire et al, 2003). When the field dominates the
pressure, it is much less affected by the gas pressure or thermodynamic
state. 

	CNM structures are magnetically subcritical. When gravitation
is important, the distinction between magnetically sub- and
supercritical clouds occurs at ${\Btot \over \Nperp} \sim 0.38$ (Nakano
\& Nakamura 1978). In subcritical clouds, the magnetic field dominates
gravity and prevents collapse. Our CNM sheets have $\Nperp \lesssim 2.6$
and the median ${\Btot}_{1/2} \sim 6.0$ $\mu$G, yielding a minimum ratio
$\sim 2.3$, far above the supercritical upper limit. Gravity is far from
important in these clouds, but if several clouds were to coalesce into a
gravitationally-important one then magnetic forces would prevent
gravitational collapse---unless   the field were destroyed in the
process, e.g.\ by the annihilation of oppositely-directed fields during
coalescence of individual clouds.  In this sense, the field is strong
and must dominate the act of star formation in denser clouds that form
from less dense interstellar gas.

\section{ASTROPHYSICAL DISCUSSION: COLUMN DENSITIES} \label{ndiscussion}

	It is well accepted that interstellar HI often lies in sheets.
Paper II showed this convincingly for the CNM structures. In the present
paper, we noted that the observed histogram of observed column density
$\Nobs$ falls monotonically with behavior close to ${\Nobs} ^{-1}$. We
then derived the pdf of the intrinsic column density for the sheets
$\Nperp$ and found that it follows equation \ref{Nobsobs} between two
limits, which are rather well defined. Equation \ref{Nobsobs} behaves
much like a ${\Nperp}^{-1}$ distribution. Figure \ref{nhstat}
convincingly shows that this is a good description of the observations.

	Suppose, first, that the CNM results from shocks. We have in
mind the McKee \& Ostriker (1977) model, in which a supernova shock
adiabatically compresses and heats the ambient gas; as the gas cools, it
does so under roughly constant pressure so its density increases, it
cools faster, and soon becomes the CNM. As the shocked gas cools it
slows. For low-velocity CNM, which is primarily what we observe, the
swept-up column density depends on the energy injected and the ambient
density. This dependence is complicated (Cioffi, McKee, \& Bertschinger
1988), so we don't attempt a detailed discussion. However, the swept-up
column densities are not incomparable with our upper limit
${\Nperp}_{max} = 2.6 \times 10^{20}$ cm$^{-2}$ in equations
\ref{Nobsobs} and \ref{Nobsobsnumbers}. Thus, the $\sim {\Nperp}^{-1}$
dependence in equation \ref{Nobsobs} could be a reflection of the
statistical distribution of the relevant function of injected energy and
ambient density. 

	Another possibility is that the CNM arises from kinematical and
thermal processes in the turbulent interstellar medium.  We have in mind
structures like those seen in numerical simulations of interstellar
turbulence such as V\'azquez-Semadeni, Gazol, \& Scalo (2000; see
references in Paper II).  Consider the ``Triad region'' discussed in
extensively in \S 8.2 of Paper II and Heiles \& Crutcher (2005).  It has
line-of-sight extent $\sim 0.05$ pc and plane-of-sky extent $\sim 20$
pc, for an aspect ratio $\sim 200$.  It also has typical turbulent
velocity $\sim 1$ km s$^{-1}$, which makes the line-of-sight crossing
time $\sim 5 \times 10^4$ yr.  This is very short---interstellar
kinematical evolution over human history! If this sheet were the result
of a slowed shock, it seems remarkable that, in the presence of ISM
density fluctuations, the distance over which the swept-up column
density is accumulated would allow the sheet to appear so coherent, and
further that it would retain its coherence to be observable.  The
alternative is, we suppose, that CNM structures are transient, and that
we can map as extensive only those that currently appear to be coherent. 

	This ${\Nperp}^{-1}$ behavior does not seem to be a capricious
result. Rather, it is a challenge to the theorists to reproduce it.
Deciding between the above two possibilities, or others, is a matter of
the explanation reproducing the ${\Nperp}^{-1}$ behavior.

\section{ASTROPHYSICAL DISCUSSION: TURBULENT VELOCITIES} \label{vdiscussion}

	In \S \ref{vaderivation} we modeled the turbulent velocities as
being anisotropic, i.e.\ perpendicular to the magnetic field, and
derived the intrinsic pdf $\phi(\Va)$ from the observed one
$\psi(\Vntobs)$. Here $\Va$ is the  the $1d$ component of turbulent
velocity; for the $2d$ case we modeled, the full component is $2^{1/2}
\Va$.  In \S \ref{bivarVBobs} we discussed the bivariate distribution
$\psi(V_{turb,los},B_{obs})$. The data fit the model with anisotropic
turbulence quite well, but the data also fit no correlation quite well,
so the results are inconclusive. The observations cannot distinguish
between anisotropic and isotropic turbulent velocities. 

	We can, however, discuss topics such as the relative energy
densities in turbulent motions and magnetism, that is, whether
turbulent velocities are superAlfv\'enic. Doing this requires some care
in definitions of parameters: we measure line-of-sight turbulent
velocity dispersions and these must be converted to their 2- or
3-dimensional counterparts; and we must include He as a component of the
ISM mass density. Finally, we can discuss results in terms of the
conventional plasma parameter $\beta$, in terms of energy densities, or
in terms of supersonic and superAlfv\'enic. 

	First we define velocities and the Mach number. Let $\Delta
V_{turb, 1d}$ be the $1d$ turbulent velocity dispersion and $\Delta
V_{th, 1d}$ be the thermal velocity dispersion. For isotropic
turbulence, the full turbulent velocity is $\Delta V_{turb} = \Delta
V_{turb, 3d} = 3^{1/2} \Delta V_{turb, 1d}$.  With this isotropic
turbulence,  the turbulent Mach number $M_{turb}$ is

\begin{equation}
M_{turb}^2 = { 3 \Delta V_{turb, 1d}^2 \over C_s^2}
\end{equation}

\noindent Here $C_s$ is the velocity of sound; the appropriate sound
velocity is the isothermal one because thermal equilibrium is reached
quickly in the CNM, so 

\begin{mathletters}
\begin{equation}
C_s^2= {nkT \over \rho}
\end{equation}

\noindent The volume density is 

\begin{equation}
\rho =(1+4f_{He}) n_H m_H
\end{equation}
\end{mathletters}

\noindent where $f_{He}$ is the fractional abundance of He by number and
$m_H$ is the mass of the H atom; we adopt $f_{He}=0.1$. Similarly, for
the Alfv\'en velocity $V_{ALF}$, we have

\begin{equation}
V_{ALF}^2= {{\Btot}^2 \over 4 \pi \rho} 
\end{equation}

\noindent The information propagation velocity perpendicular to the
field lines is equal to the Alfv\'en velocity $V_{ALF}$. The mean square
velocity perpendicular to the field lines is twice the line-of-sight
value. Consequently, we define the Alfv\'enic turbulent Mach number
$M_{ALF,turb}$ as

\begin{equation}
M_{ALF,turb}^2= {2 \Delta V_{turb,1d}^2 \over V_{ALF}^2} 
\end{equation}

\noindent If $M_{ALF,turb} > 1$, then then shocks will develop; this is
the superAlfv\'enic case. 

	Next we define energy densities. The turbulent energy density is

\begin{equation}
E_{turb} = {  \rho F \Delta V_{turb, 1d}^2 \over 2}
\end{equation}

\noindent Here we include the quantity $F$ to allow for anisotropic
turbulence. If turbulent velocities are only perpendicular to $B$, then
$F=2$; isotropic turbulence has $F=3$ (the case assumed in paper II). 
Of course, the magnetic energy density is 

\begin{equation}
E_{mag}= {{\Btot}^2 \over 8 \pi}
\end{equation}

\noindent The ratio is

\begin{equation}
{ E_{turb} \over E_{mag}} = {F \Delta V_{turb, 1d}^2 \over V_{ALF}^2 } 
	= {F \over 2} M^2_{ALF, turb} \ .
\end{equation}

	Finally, we define the conventional plasma parameter
$\beta_{th}$, which compares thermal and magnetic pressures:

\begin{equation}
\beta_{th} \equiv {P_{th} \over P_{mag}} = 
	{2 \Delta V_{th,1d}^2 \over V_{ALF}^2} { 1+f_{He} \over 1+4f_{He}}
\end{equation}

	For comparison of turbulent and magnetic effects, we calculate
the relevant ratios for the following adopted parameter values, which
are close to the medians:

\begin{mathletters} \label{typicalvalues}
\begin{equation} 
T_{CNM} = 50 \ {\rm K} \ ;
\end{equation}
\begin{equation}
{P_{CNM} \over k}= 3000 \ {\rm cm^{-3} \ K} \ ;
\end{equation}
\begin{equation}
n(HI)_{CNM} = 54 \ {\rm cm}^{-3} \ ;
\end{equation}
\begin{equation}
\Delta V_{turb,1d} = 1.2 \ {\rm km \ s}^{-1} \ ;
\end{equation}
\begin{equation}
\Btot = 6.0 \ \mu{\rm G} \ .
\end{equation}
\end{mathletters}

\noindent Here $T$ is a typical CNM temperature from Paper II. 
$P_{CNM}$ is from Jenkins \& Tripp (2001) and Wolfire et al.\ (2003).
The value for $\Delta V_{turb,1d}=1.2$ km s$^{-1}$ is the median from \S
\ref{vaderivation}. The value for $\Btot$ is the median from \S
\ref{montecarlo}. 

	These values provide

\begin{mathletters}
\begin{equation}
M_{turb} = 3.7 \ ;
\end{equation}
\begin{equation}
\beta_{th} = 0.29  \ ;
\end{equation}
\begin{equation}
M^2_{ALF,turb} = 1.3  \ ;
\end{equation}
\begin{equation}
V_{ALF} = 1.5  \ {\rm km \ s}^{-1} \ .
\end{equation}
\end{mathletters}

\noindent 
	If the field were small, then turbulence would be isotropic with
$F=3$. However, it is not so small, but neither is it so large that we
could say for certain that $F=2$.  The limits $2 < F < 3$ correspond to 
$1.3 < {E_{turb} \over E_{mag}} < 1.9$.

	These values should be regarded as representative. Not all CNM
clouds have the median values, so these parameters have a considerable
spread.

	One interesting question is whether {\it any} CNM Clouds have
high $\beta_{th}$, i.e.\ whether any CNM clouds have negligible magnetic
field. We cannot answer this because $\beta_{th} \propto {P \over
{\Btot}^2}$ and neither $P$ nor $\Btot^2$ is attainable for an
individual cloud. Moreover, the functional form of $B_{tot}$ is ill
determined: we cannot even distinguish between all values being the same
(a delta-function distribution) and a Gaussian (which has a significant
population of very high field strengths); see \S \ref{montecarlo}.

\section{FINAL COMMENTS: TURBULENT AND MAGNETIC ENERGY EQUIPARTITION}

\label{finalcomments}

      Our numbers indicate that magnetism and turbulence are in
approximate equipartition.  The approximate equipartition suggests that
turbulence and magnetism are intimately related by mutual exchange of
energy.  In the absence of turbulence, magnetic energies do not
dissipate because the magnetic field cannot decay on short time scales. 
But with turbulence, the field may be able to decay rapidly (Heitsch \&
Zweibel 2003; Lazarian \& Visniac 1999; Zweibel 2002).  Moreover,
numerical simulations suggest that supersonic turbulence also dissipates
rapidly, even when the field is strong (MacLow et al 1998).  However, it
is not obvious to us that these dissipative processes should lead to the
observed equipartition between turbulence and magnetism. 

        We suspect the answer lies in Hennebelle \& Perault's (1999)
result and Maclow's (private communication) observation that the CNM
components result from the transient nature of turbulent flow: the CNM
occupies regions where densities are high, produced by converging flows,
and the density rise is limited by pressure forces.  These limiting
pressures are magnetic because the gas has small $\beta_{th}$, meaning
that thermal pressure is negligible and the dynamical equality makes the
magnetic pressure comparable to the converging ram pressure and leads to
apparent equiparition.  

	The equipartition looks like a steady-state equilibrium, but it
is really a snapshot of time varying density fields and our immediate
observational view is a statistical result over a large sample.  In
other words, our current observational snapshot shows an ensemble at a
given time.  Against this we compare the numerical simulations, which
are stationary in the sense that they have been allowed to run long
enough that the statistical properties become time-independent.  Such
simulations are also ergodic, with statistical properties over time
being equivalent to those over space.  With this view, the ISM
dynamically evolves through turbulence and its properties are governed
by statistical equilibrium of energy inputs and dissipation.  These
matters are discussed at length in the excellent review by MacLow \&
Klessen (2004)

\acknowledgements

	It is a pleasure to acknowledge discussions with Martin Bureau,
Jeremy Darling, Avinash Deshpande, Doug Finkbeiner, Patrick Hennebelle,
Dave Hollenbach, Mordecai MacLow, and Chris McKee.  This work was
supported in part by NSF grants AST-9530590, AST-0097417, AST-9988341,
and AST 04-06987; and by the NAIC.

\end{document}